\RequirePackage{fix-cm}
\documentclass[twocolumn,epjc3,envcountsect]{svjour3}
\RequirePackage{graphicx}
\RequirePackage{mathptmx}      
\RequirePackage{flushend}
\RequirePackage[colorlinks,citecolor=blue,urlcolor=blue,linkcolor=blue]{hyperref}
\makeatletter
\renewcommand{\makeheadbox}{}
\makeatother

\usepackage{pdfpages}
\usepackage{hyperref}
\usepackage{orcidlink}
\usepackage{xcolor}
\usepackage{physics}
\usepackage{xfrac}

\usepackage[utf8]{inputenc}

\usepackage{csquotes}
\usepackage[english]{babel}
\usepackage{subfigure}
\usepackage[parfill]{parskip} 

\usepackage[
  backend=biber,
  bibstyle=numeric-epj,
  citestyle=numeric-comp,
  alldates=iso,
  seconds=true,
  doi=true,
  url=false,
  isbn=false,
  eprint=true,
  giveninits=true,
]{biblatex}
\DeclareFieldFormat{pages}{#1}

\renewbibmacro{in:}{}

\renewbibmacro*{doi+eprint+url}{%
    \printfield{doi}%
    \newunit\newblock%
    \iffieldundef{doi}{%
    \iftoggle{bbx:eprint}{%
        \usebibmacro{eprint}%
    }{}%
    }{}%
    \newunit\newblock%
    \iffieldundef{doi}{%
        \usebibmacro{url+urldate}}%
        {}%
    }

\bibliography{abbrv_ltwa,bib}

\usepackage{textgreek}
\usepackage[
retain-explicit-plus,
uncertainty-mode = separate,
separate-uncertainty-units = single
]{siunitx}
\usepackage{xspace}

\DeclareSIUnit\barn{b}

\newcommand{\eq}{\begin{equation}}
\newcommand{\eeq}{\end{equation}}

\newcommand{\decays}{$\to$}

\newcommand{\prot}{p\xspace}

\newcommand{\pprot}{\prot{}$_\text{p}$\xspace}
\newcommand{\deut}{d\xspace}
\newcommand{\neut}{n\xspace}
\newcommand{\nucl}{N\xspace}
\newcommand{\hype}{Y\xspace}
\newcommand{\hyped}{Y\xspace}
\newcommand{\rest}{X\xspace}

\newcommand{\nstar}{\nucl{}$^{*}$\xspace}

\newcommand{\elec}{e\xspace}
\newcommand{\elp}{\elec{$^{+}$}\xspace}
\newcommand{\elm}{\elec{$^{-}$}\xspace}
\newcommand{\dalitzp}{\elp\elm}

\newcommand{\rg}{\textgamma\xspace}
\newcommand{\vg}{\rg{$^{*}$}\xspace}

\newcommand{\mpi}{\textpi\xspace}

\newcommand{\piz}{\mpi{$^{0}$}\xspace}
\newcommand{\pim}{\mpi{$^{-}$}\xspace}
\newcommand{\pip}{\mpi{$^{+}$}\xspace}

\newcommand{\Kaon}{K\xspace}
\newcommand{\aKaon}{$\overline{\mbox{\Kaon}}$\xspace}

\newcommand{\Kz}{\Kaon{$^{0}$}\xspace}
\newcommand{\Kzs}{\Kaon{$^{0}_\text{S}$}\xspace}
\newcommand{\aKzs}{\aKaon{$^{0}_\text{S}$}\xspace}

\newcommand{\Km}{\Kaon{$^{-}$}\xspace}
\newcommand{\Kp}{\Kaon{$^{+}$}\xspace}

\newcommand{\Dz}{\textDelta{$^{0}$}\xspace}

\newcommand{\Dpp}{\textDelta{$^{++}$}\xspace}

\newcommand{\Shype}{\textSigma\xspace}  
\newcommand{\Szero}{\Shype{$^0$}\xspace}

\newcommand{\Splus}{\Shype{$^{+}$}\xspace}

\newcommand{\Lhype}{\textLambda\xspace}
\newcommand{\Lzero}{\Lhype{}\xspace}

\newcommand{\Xihype}{\textXi\xspace}

\newcommand{\Xim}{\Xihype{}$^{-}$\xspace}

\newcommand{\dilambda}{\Lzero\Lzero}

\newcommand{\pp}{\prot{}+\prot{}\xspace}

\newcommand{\pimp}{\pim+\prot\xspace}

\xspaceaddexceptions{\decays \Kaon \aKaon \prot \pprot \deut \neut \nucl \hype \hyped \rest \nstar \elec \vg \photon \elp \elm \piz \pim \pip \Kzs \Kz \aKzs \Km \Kp \Dz \Dpp \Shype \Szero \Splus \Sstar \starA \Sstarz \Sstarp \Lzero \Lstar \Lstard \Xim \pp \pimp \dilambda \dalitzp + \textgamma \textpi}

\DeclareSIUnit\clight{\text{\ensuremath{c}}}
\DeclareSIUnit\mclight{c}

\DeclareSIUnit\mub{\micro\barn}

\DeclareSIUnit\tev{\tera\electronvolt}

\DeclareSIUnit\gev{\giga\electronvolt}
\DeclareSIUnit\gevc{\giga\electronvolt\per\clight}
\DeclareSIUnit\gevsc{\giga\electronvolt\per\square\clight}
\DeclareSIUnit\sgevsc{\giga\electronvolt\squared\per\clight\tothe{4}}
\DeclareSIUnit[mode=text]{\sgevc}{(GeV/c)^{2}}
\DeclareSIUnit{\sgevc}{(\giga\electronvolt\per\clight)\squared}
\DeclareSIUnit{\sgevcinv}{(\clight\per\giga\electronvolt)\squared}

\DeclareSIUnit\mev{\mega\electronvolt}
\DeclareSIUnit\mevc{\mega\electronvolt\per\clight}
\DeclareSIUnit\mevsc{\mega\electronvolt\per\clight\squared}
\DeclareSIUnit\smevsc{\mega\electronvolt\squared\per\clight\tothe{4}}
\DeclareSIUnit\mevs{\mega\electronvolt\squared}

\DeclareSIUnit\kev{\kilo\electronvolt}
\DeclareSIUnit\kevc{\kilo\electronvolt\per\clight}
\DeclareSIUnit\kevsc{\kilo\electronvolt\per\clight\squared}

\DeclareSIUnit\bar{bar}

\newcommand{\brange}[4][]{%
\SIrange[range-units=bracket,range-open-bracket={[},range-phrase={ , },range-close-bracket={]},#1]{#2}{#3}{#4}%
}

\usepackage{tabularx}
\usepackage{etoolbox}
\usepackage{booktabs}
\usepackage{multicol}
\usepackage{xcolor}

\robustify\bfseries
\newrobustcmd{\B}{\bfseries}
\newcolumntype{Y}{>{\centering\arraybackslash}X}

\newcolumntype{R}{>{\raggedleft\arraybackslash}X}

\usepackage[capitalize]{cleveref}

\begin{document}

\title{Measurement of Differential Cross Sections and Integrated Luminosity using Proton-Proton Elastic Scattering with HADES at 
\textit{T} = 4.53 GeV and \textit{T} = 1.60 GeV
}

\author{HADES collaboration \\[5bp]
R.~Abou~Yassine$^{7,14}$, J.~Adamczewski-Musch\orcidlink{0000-0002-7586-8504}$^{6}$, G.~A.~Appagere$^{16,*}$, M.~Becker$^{11}$,
A.~Blanco\orcidlink{0000-0001-9827-8294}$^{2}$, C.~Blume\orcidlink{0000-0002-6800-3465}$^{9,6,f}$, M.~Bohman\orcidlink{0009-0007-6118-8683}$^{17,*}$, Y.~Bondar\orcidlink{0000-0003-2773-9668}$^{4}$, L.~Chlad\orcidlink{0000-0003-3814-2920}$^{15,h}$,
I.~Ciepa{\l}\orcidlink{0000-0003-0936-3054}$^{4}$
% ...
}

\institute{
\mbox{} \\[-8bp]
\mbox{$^{1}$Ruhr-Universit\"{a}t Bochum, 44801~Bochum, Germany}\\
\mbox{$^{2}$LIP-Laborat\'{o}rio de Instrumenta\c{c}\~{a}o e F\'{\i}sica Experimental de Part\'{\i}culas, 3004-516~Coimbra, Portugal}\\
\mbox{$^{3}$AGH University of Krakow, Faculty of Physics and Applied Computer Science, 30-059~Krakow, Poland}\\
\mbox{$^{4}$Institute of Nuclear Physics, Polish Academy of Sciences, 31342~Krak\'{o}w, Poland}\\
\mbox{$^{5}$Smoluchowski Institute of Physics, Jagiellonian University of Cracow, 30-059~Krak\'{o}w, Poland}\\
\mbox{$^{6}$GSI Helmholtzzentrum f\"{u}r Schwerionenforschung GmbH, 64291~Darmstadt, Germany}\\
\mbox{$^{7}$Institut f\"{u}r Kernphysik, Technische Universit\"{a}t Darmstadt, 64289~Darmstadt, Germany}\\
\mbox{$^{8}$Institut f\"{u}r Strahlenphysik, Helmholtz-Zentrum Dresden-Rossendorf, 01314~Dresden, Germany}\\
\mbox{$^{9}$Institut f\"{u}r Kernphysik, Goethe-Universit\"{a}t, 60438 ~Frankfurt, Germany}\\
\mbox{$^{10}$Physik Department E62, Technische Universit\"{a}t M\"{u}nchen, 85748~Garching, Germany}\\
\mbox{$^{11}$II.Physikalisches Institut, Justus Liebig Universit\"{a}t Giessen, 35392~Giessen, Germany}\\
\mbox{$^{12}$Department of Mechanical Engineering, Frederick University, 1036~Nicosia, Cyprus}\\
\mbox{$^{13}$Faculty of Science, Palack\'{y} University Olomouc, 779 00~Olomouc, Czech Republic}\\
\mbox{$^{14}$Laboratoire de Physique des 2 infinis Irene Joliot-Curie, Universite Paris-Saclay, CNRS-IN2P3, F-91405~Orsay, France}\\
\mbox{$^{15}$Nuclear Physics Institute, The Czech Academy of Sciences, 25068~Rez, Czech Republic}\\
\mbox{$^{16}$Department of Physics, Stockholm University, ~Stockholm, Sweden}\\
\mbox{$^{17}$Institutionen for fysik och astronomi, Uppsala universitet, 75120~Uppsala, Sweden}\\
\mbox{$^{18}$Uniwersytet Warszawski, Instytut Fizyki Do\'{s}wiadczalnej, 02-093~Warszawa, Poland}\\
\mbox{$^{19}$Warsaw University of Technology, Faculty of Physics, 00-662~Warsaw, Poland}\\
\mbox{$^{20}$Bergische Universit\"{a}t Wuppertal, 42119~Wuppertal, Germany}\\
\\
\mbox{$^{*}$ members of the PANDA@HADES collaboration}\\
\mbox{$^{a}$ also at Helmholtz Research Academy Hesse for FAIR (HFHF), Campus Giessen, ~Giessen, Giessen}\\
\mbox{$^{b}$ also at Instituto Politecnico de Coimbra, Instituto Superior de Engenharia de Coimbra, 3030-199~Coimbra, Portugal}\\
\mbox{$^{c}$ also at Doctoral School of Exact and Natural Sciences, Jagiellonian University, ~Cracow, Poland}\\
\mbox{$^{d}$ also at Helmholtz Research Academy Hesse for FAIR (HFHF), Campus Darmstadt, 64390~Darmstadt, Germany}\\
\mbox{$^{e}$ also at Technische Universit\"{a}t Dresden, 01062~Dresden, Germany}\\
\mbox{$^{f}$ also at Helmholtz Research Academy Hesse for FAIR (HFHF), Campus Frankfurt, 60438~Frankfurt am Main, Germany}\\
\mbox{$^{g}$ also at Charles University, Faculty of Mathematics and Physics, 12116~Prague, Czech Republic}\\
\mbox{$^{h}$ also at Czech Technical University in Prague, 16000~Prague, Czech Republic}\\
\mbox{ e-mail: hades-info@gsi.de (J.~Stroth)}\\
}

\authorrunning{HADES Collaboration}
\titlerunning{Measurement of $pp\rightarrow pp$ Cross Sections and Integrated Luminosity} 
\date{Received: date / Revised version: date}

\onecolumn
\maketitle

\begin{multicols}{2}

\abstract{This work presents an investigation of elastic proton-proton scattering based on data collected with the recently upgraded HADES detector using beams of protons with kinetic energies of $T = \SI{4.53}{\gev}$ and  $T = \SI{1.60}{\gev}$ impinging on a liquid hydrogen target.
This is the first measurement emerging from the upgraded detector setup, including a Forward Detector with straw-tube tracking planes designed for the PANDA experiment. The well-constrained two-body kinematic relations of elastic scattering have been exploited to study the detector alignment, the beam alignment, and to provide the final measurement of the beam energy. In addition, the recorded luminosity of the experiment has been determined by normalizing the HADES data in a range of the squared four-momentum transfer $t$ that overlaps with measurements by previous experiments worldwide. Furthermore, the differential cross sections have been determined over a large range of $t$ from which the slope parameter $B$ is extracted, yielding $B = \SI[per-mode=power]{8.98 \pm 0.3}{(\gevc)^{-2}}$ and $\SI[per-mode=power]{7.3 \pm 0.1}{(\gevc)^{-2}}$ at $T = \SI{4.53}{\gev}$ and $\SI{1.60}{\gev}$, respectively.
This contribution to the world database on elastic proton-proton scattering provides input to theoretical modeling and serves as a reference for luminosity determinations by future experiments in the same energy range.
\PACS{
      {25.40.Cm} {Elastic proton scattering} \and
      {13.75.Cs} {Nucleon-nucleon interactions} \and
      {13.85.Dz} {Elastic scattering}
     }
}
\end{multicols}
\twocolumn

\section{Introduction}
\label{sec:introduction}

The HADES experiment at GSI, Germany, has recently undergone an upgrade that facilitates an ambitious hyperon physics program using proton-proton collisions \cite{HADES:2020pcx}. For normalization of the data and the subsequent cross-section measurement, a precise determination of the time-integrated luminosity is essential. Elastic proton-proton scattering is ideal for this purpose by virtue of its two-body kinematics, which gives rise to a distinct signature in most particle detectors. Partial Wave Analysis models \cite{Workman:2016ysf} have been used to interpret experimental data and enabled a precise and coherent description of scattering processes at low energies. Existing data and partial wave fits serve as a reference for luminosity determinations by new experiments. 
In addition, the distinct kinematic signature of elastic scattering provides an ideal tool to measure the energy of the beam and its angle of incidence on the target. Furthermore, it is an excellent reaction to use for controlling the detector alignment and the momentum reconstruction.

Elastic proton-proton scattering is also of more general interest, as it is important for understanding the strong interaction at all energy scales. At very low energies, below the pion-production threshold, the wealth of nucleon-nucleon scattering data provides input to \textit{ab initio} calculations of nuclear properties using \textit{e.g.} Effective Field Theories \cite{Machleidt:2024bwl}. At higher energies, from slightly below 1 \unit{\gev} up to a few \unit{\gev}, elastic scattering amplitudes can reveal resonances \cite{Arndt:1997if}, constrain models of nuclear matter \cite{Shirokov} or be used as input to calculations of more complex many-body reactions \cite{Dymov:2015hku}. At the \unit{\gev} to \unit{\tev} scale, the elastic cross-section dependence on the momentum-transfer squared $t$ probes the nucleon structure \cite{Islam:2005by} and offers a testing ground for Pomeron and Odderon models~\cite{Selyugin:2012gv}. 

The relevant kinematic variables in unpolarized proton-proton elastic scattering are the beam kinetic energy $T$ and the four-momentum transfer $t$. The interaction dynamics manifest in the scattering amplitude and hence the differential cross section. At $|t|> \SI{0.01}{{\sgevc}}$, the differential cross section $\sigma_\mathrm{tot}$ is dominated by the nuclear amplitude $\sigma_n$, which can be parameterized in terms of the slope parameter $B$ and the parameter $C$ that encodes non-linearities of the slope of the exponent at large $|t|$ \cite{Uzhinsky:2016uhg}:
\begin{align}
	\dv{\sigma_n}{t}  
 = \dv{\sigma_n}{t}\bigg|_{t=0}e^{-B|t|+Ct^2}. \label{eq:quadratic_exponential}
\end{align}
For $|t|<\SI{0.1}{{\sgevc}}$ the $Ct^2$ term in the exponential function can be considered negligible and the exponential therefore simplifies to
\begin{equation}
    \dv{\sigma_n}{t} = \dv{\sigma_n}{t}\bigg|_{t=0}e^{-B|t|}. \label{eq:linear_exponential}
\end{equation}
The value of the nuclear differential cross section at $t=0$, \textit{i.e.} $\dv{\sigma_n}{t}|_{t=0}$ is called the optical point. 

In this work, elastic proton-proton scattering has been studied with the HADES detector with a liquid hydrogen target and a proton beam with kinetic energies $T = \SI{4.53}{\gev}$ and $T = \SI{1.60}{\gev}$, corresponding to beam momenta of $p_{\mathrm{beam}} = \SI{5.39}{\gevc}$ and $p_{\mathrm{beam}} = \SI{2.36}{\gevc}$, respectively. However, in most of this paper, the beam kinetic energy is used when referring to these data. The differential cross section has been determined with various detector configurations in the kinematic ranges of $\SI{-4.13}{{\sgevc}}< t < \SI{-0.1}{{\sgevc}}$ and $\SI{-1.44}{{\sgevc}}< t < \SI{-0.14}{{\sgevc}}$ for the data at $T = \SI{4.53}{\gev}$ and $T = \SI{1.60}{\gev}$, respectively.

The data were exploited in a two-fold endeavor. First, the yield of elastic proton-proton scattering events was extracted in a region of $t$ in which cross sections have already been measured by other experiments. This provided the necessary information for a precise determination of the luminosity recorded by HADES. Next, $t$-regions that were previously less known or even unknown were studied to extract differential cross sections. In this way, these new data contribute to the world database and thereby to our knowledge of the mechanism of hadron scattering, as well as to future experiments that may use these results. 

Elastic nucleon-nucleon scattering measurements are compiled in the SAID database \cite{SAID}. SAID also provides partial wave analysis solutions for proton-proton scattering from threshold up to a beam kinetic energy of ${T = \SI{3}{\gev}}$~\cite{Workman:2016ysf}. Moreover, high-precision data have been provided by EDDA at COSY for $\theta_\mathrm{CM}>\ang{35}$ ($t=\SI{-0.27}{{\sgevc}}$) within the same energy range~\cite{EPJA.22.125_COSY_low_energy}. The resulting fits describe well the available differential cross-section data and have been used for the normalization of the HADES data at $T = \SI{1.60}{\gev}$. At this beam energy, HADES covers the $t$ region up to $t = \SI{-0.14}{{\sgevc}}$, for which no measurement previously existed.

At $T = \SI{4.53}{\gev}$, the most precise cross sections in this energy region are measured at Argonne and at the Bevatron at small momentum transfer $\num{-0.66} \leq t \leq \SI{-0.05}{{\sgevc}}$ ~\cite{PhysRev.170.1223, PhysRevD.9.1179, PhysRevD.4.1309, Clyde:1966rta, PhysRev.107.859, PhysRev.154.1284}. In $t$-regions where the HADES acceptance overlaps with that of other experiments at similar beam energies, previously measured cross sections have been interpolated to the HADES energy and used as reference data for the determination of the recorded time-integrated luminosity. By virtue of the large available data sample, HADES can provide precise data especially at large scattering angles. At this energy, no previous measurements exist and the strong energy dependence of the cross section makes interpolation between energies unreliable. With the data collected by HADES at $T = \SI{4.53}{\gev}$, differential cross sections and the slope parameter $B$ have been extracted at a new energy point. Precise measurements of these quantities over a wide energy range provide essential input for advancing nuclear and hadron physics toward higher precision.

This paper is organized as follows: In \cref{sec:experimentalsetup}, the HADES experiment is presented with emphasis on the recent upgrade and the data collection campaign from 2022 with a proton beam. This is followed by a brief presentation of the data sample in \cref{sec:datasamples} and a description of the two-body kinematic relations that characterize elastic scattering in \cref{sec:kinrel}. In \cref{sec:beamtargetconditions},  an investigation of the detector and beam alignment and an independent measurement of the beam energy is presented. The event selection used in the physics analysis is outlined in \cref{sec:eventselection} and the resulting acceptance and efficiency is presented in \cref{sec:acceptanceefficiency}. In \cref{sec:systematics}, an investigation of systematic uncertainties is presented, and in \cref{sec:luminosity} the determination of the recorded time-integrated luminosity of this data campaign. The resulting $t$-dependent differential cross sections are presented in \cref{sec:crosssectionmeasurements} and  finally a summary is given in \cref{sec:discussion}.

\section{Experimental Setup}
\label{sec:experimentalsetup}

\begin{figure}[!t]
  \centering
  \includegraphics[width=1.0\linewidth]{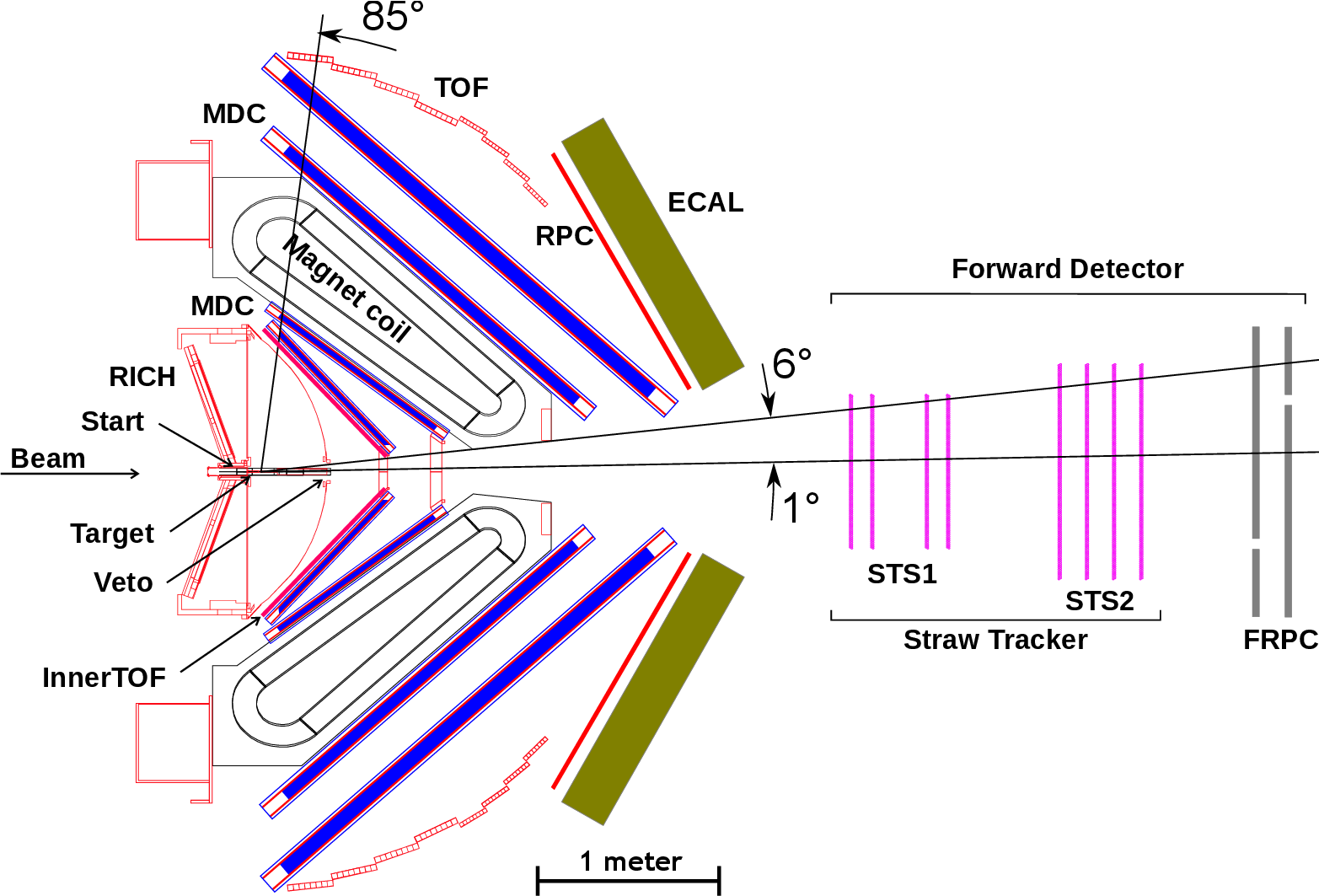}
  \caption{Cross-sectional view of the upgraded HADES spectrometer, including the new Forward Detector.}
  \label{fig:hades-feb22}
\end{figure}

This work presents the first results from the upgraded setup of the High Acceptance Di-Electron Spectrometer (HADES, see Ref.~\cite{Agakichiev_2009_HADES}). The proton beam for the experiment was delivered by the Heavy-Ion Synchrotron (SIS18) at GSI, and impinged on a liquid hydrogen target. The setup of the spectrometer for this experiment is shown in \cref{fig:hades-feb22}. 
The main HADES spectrometer has been extended by a Forward Detector~\cite{HADES:2020pcx} using straw tubes \cite{Smyrski:2017,Smyrski:2018} designed for the PANDA experiment \cite{PANDA:2021ozp}, and Resistive Plate Chambers (RPC) for Time-of-Flight (ToF) measurements. In the following, the HADES experimental setup with its main components is described: the target, the main HADES spectrometer, and the Forward Detector.

\subsection{Target}
The liquid hydrogen target with areal density \SI{2e23}{cm^{-2}} is contained within a \SI{5}{\cm} long cylinder with a diameter of \SI{2.5}{\cm} operated at atmospheric pressure and \SI{20}{\kelvin}. The target volume is surrounded by a \SI{100}{\um} Mylar foil, with an additional \SI{100}{\um} foil layer located \SI{6}{\mm} downstream of the target \cite{Agakichiev_2009_HADES}. The proton beam has a \SI{0.7}{\percent} nuclear interaction probability in the target material. The threshold kinetic energy to leave the target and its material is $\SI{15}{\mev}$ for protons emitted at a polar angle of \ang{90}.

\subsection{Main HADES Spectrometer}
\label{subsec:mainHADES}
The main HADES spectrometer consists of six sectors covering \SI{85}{\percent} of the azimuthal angle over a polar angle interval from \ang{18} to \ang{85}. The event start time is provided by the START detector~\cite{KRUGER2022167046}, based on the Low Gain Avalanche Diode (LGAD) technology and placed upstream of the liquid hydrogen target. Charged particles are tracked by four planes of Mini Drift Chambers (MDC), which are placed pairwise in the magnetic field-free region before and after the magnet, respectively. A toroidal magnetic field is confined to the space between MDC planes II and III, facilitating momentum determination. 
A plastic-scintillator-based Time-of-Flight (TOF) system detects particles emitted at polar angles ${\ang{44} \leq \vartheta \leq \ang{85}}$, while the RPC planes cover ${\ang{18} \leq \vartheta \leq \ang{44}}$. In addition to the outer TOF detectors, an inner Time-of-Flight detector (iTOF or InnerTOF), consisting of plastic scintillator plates~\cite{GRZONKA2022167410}, is placed just in front of the innermost MDC plane and covers the full HADES acceptance. More details of the experimental setup can be found in Ref.~\cite{Agakichiev_2009_HADES}.

\subsection{Forward Detector}
\label{subsec:forwarddetector}

The Forward Detector (FD) system was installed in 2020. It consists of two Straw Tube Stations (STS) for charged particle track reconstruction and a Forward Resistive Plate Chamber (FRPC) for ToF measurements. The FD system increases the polar angle acceptance of the HADES spectrometer to include the detection of charged particles emitted in the region $\ang{1} < \vartheta < \ang{6}$. More details of the Forward Detector than included below can be found in Ref.~\cite{HADES:2020pcx_forward_detector}. 

\subsubsection*{Straw Tube Stations (STS)}

The straw tube tracking stations are based on self-supporting straw tube detectors with \SI{10}{\mm} inner diameter with a \SI{27}{\um} thick Mylar cathode wall~\cite{Wintz:2014uwa,Smyrski:2017}. The individual straw tubes are glued together into double-layers, and four such double-layers are combined to make each of the two stations STS1 and STS2 located \SI{3.1}{\metre} and \SI{4.6}{\metre} downstream of the target, respectively (see \cref{fig:hades-feb22}). The straw tubes in STS1 and STS2 have lengths of \SI{76}{\cm} and \SI{125}{\cm}, respectively. The double-layers are arranged normal to the beam ($z$-axis) and are rotated azimuthally by \ang{0}, \ang{90}, \ang{90}, \ang{0} and \ang{0}, \ang{90}, \ang{-45}, \ang{+45} for STS1 and STS2, respectively. The central tubes are shorter to allow the beam to pass. 

The tubes are filled with an Ar:CO$_2$ (90:10) mixture and operated at \SI{1}{\bar} pressure above the ambient atmospheric pressure. This overpressure provides the mechanical stiffness required for the straw tubes, thereby making them self-supporting. The spatial resolution for each straw tube layer has been measured to be about \SI{0.13}{\mm} ($\sigma$) for minimum-ionizing protons~\cite{Smyrski:2018}. 

\subsubsection*{Forward Resistive Plate Chambers (FRPC)}
The FRPC is a ToF system based on individually shielded resistive plate chambers~\cite{FINCK200363}. Each module consists of 32 RPC counters of length \SI{750}{\mm} and two different widths, arranged in partially overlapping double layers ($2 \times 16$ counters: 20 of \SI{22}{\mm} (inner side) followed by 12 of \SI{42}{\mm} (outer side)). Four of these modules, rotated and aligned such that the inner side is always at smaller polar angles, are used to make full azimuthal coverage of the STS tracking acceptance region (see \cref{fig:hades-feb22}). The FRPC is located \SI{5.5}{\metre} downstream of the target.
The detector operates at particle fluxes of \SI{320}{particles s{^-1}\cm^{-2}} with an overall detection efficiency of $\approx\SI{91}{\percent}$ and a time resolution of $\approx$ \SI{112}{\pico\second} \cite{BLANCO2023167652}.

\subsection{Trigger}
\label{subsubsec:trigger}

The elastic scattering data were recorded with a minimum bias trigger (PT2), which accepts events with at least one hit in the iTOF and one hit in either the RPC or TOF of the same sector. This trigger is down-scaled by a factor of 64. The average trigger rate during the beamtime was 26 kHz, including the beam duty factor which was found to be about 55\,\% due to the synchrotron spill structure. The trigger rate was dominated by the main physics trigger (PT3) that selects events with at least two iTOF and RPC/TOF coincidences plus at least one additional hit in the iTOF. 

\section{Data Samples}
\label{sec:datasamples}

The data presented here were recorded during February and March 2022. The data from this campaign are denoted Feb22 throughout this paper. The beamtime was divided into five periods, denoted I, IIA, IIB, III and IV. The data from period I, IIA, IIB and III were collected at a beam kinetic energy of $T = \SI{4.53}{\gev}$ and the data from period IV at $T = \SI{1.60}{\gev}$.

The efficiency and acceptance were studied using Monte Carlo (MC) simulations of elastic scattering events with realistic angular distributions using the Pluto event generator~\cite{pluto}. The generated elastically scattered protons were propagated through the detector using Geant3 and the HADES reconstruction software HYDRA. The angular distribution that was used as input to the MC generator at $T = \SI{4.53}{\gev}$ data was obtained by parameterizing interpolated data between existing data at near-lying energies. To generate the corresponding distribution at $T = \SI{1.60}{\gev}$, we utilized a parameterization from a SAID PWA analysis, integrated into Pluto. A total of \num{5e8} elastic scattering events were simulated at $T = \SI{4.53}{\gev}$ and \num{5e7} events at $T = \SI{1.60}{\gev}$. This corresponds to about half of the number of signal events recorded during the experiment (see \cref{tab:el_cut_flow_hh,tab:el_cut_flow_h,tab:el_cut_flow_hf}).

The data were divided into three distinct categories according to which part of the detector the scattered protons traverse, corresponding to complementary regions in $t$: 

\begin{enumerate}
 \item HH (HADES-HADES) -- both protons are detected in the main \mbox{HADES} spectrometer:
 \begin{itemize}
  \item $T = \SI{4.53}{\gev}$: for cross-section measurements in the region of $\SI{-4.13}{{\sgevc}} < t < \SI{-2.03}{{\sgevc}}$.
  \item $T = \SI{1.60}{\gev}$: for luminosity measurement in the region of $\SI{-1.44}{{\sgevc}} < t < \SI{-0.44}{{\sgevc}}$.
 \end{itemize}
 \item HS (HADES-Single) -- one proton in the main HADES spectrometer and the other undetected:
 \begin{itemize}
  \item $T = \SI{1.60}{\gev}$: for cross-section measurements in the region of $\SI{-0.66}{{\sgevc}} < t < \SI{-0.14}{{\sgevc}}$.
 \end{itemize}
 \item HF (HADES-Forward) -- one proton in the main HADES spectrometer and the other in the Forward Detector:
 \begin{itemize}
  \item $T = \SI{4.53}{\gev}$: for luminosity measurement in the region of $\SI{-0.2}{{\sgevc}} < t < \SI{-0.1}{{\sgevc}}$.
 \end{itemize}
\end{enumerate}

Note that for those events at $T = \SI{1.60}{\gev}$ in which the scattered proton entered the Forward Detector, the recoil proton was emitted outside of the HADES acceptance. Therefore, no HF-coincidence sample was recorded at this energy.

\section{Kinematic Relations in Elastic Scattering}
\label{sec:kinrel}

When selecting elastically scattered events, we exploit the well-defined kinematics of a two-body reaction. For an ideal detector with perfect resolution, a set of relations are identified that are specific for elastic scattering. Firstly, in the center-of-mass (CM) system the final state protons are emitted back-to-back with momenta of equal magnitude. In the laboratory system, their motion normal to the beam direction is in the same plane, hence they are \textit{coplanar}. Denoting the azimuthal angles of the two final-state protons $\varphi_{1,2}$, the coplanarity relation can be expressed as follows
\begin{equation}
  \Delta \varphi = |\varphi_{1}-\varphi_{2}| - \ang{180} = 0
  \label{eq:coplanarity}.
\end{equation}
Secondly, the polar angles in the laboratory system $\vartheta_{i}$ are related to the Lorentz factor of the beam-target system $\gamma_{\mathrm{CM}}$:
\begin{equation}
  \Delta (\tan\vartheta_{1}\tan\vartheta_{2}) =
    \tan\vartheta_{1}\tan\vartheta_{2} - \frac{1}{\gamma^2_\mathrm{CM}} = 0
  \label{eq:tanthetaproduct}.
\end{equation}
Finally, the momentum of the elastically scattered proton in the laboratory depends on the scattering angle as
\begin{equation}
  p_\mathrm{theo} = \frac{p_\mathrm{beam}}{\cos\vartheta_{i}\qty[1+\gamma^2_\mathrm{CM}\tan^2\vartheta_{i}]}
  \label{eq:el_poftheta},
\end{equation}
where $p_\mathrm{beam}$ is the beam momentum. This yields the third elastic scattering relation:
\begin{equation}
  \Delta p = p_\mathrm{reco} - p_\mathrm{theo} = 0
  \label{eq:deltap},
\end{equation}
where $p_\mathrm{reco}$ is the momentum reconstructed by the detector. Note that the relations \cref{eq:coplanarity,eq:tanthetaproduct,eq:el_poftheta,eq:deltap} are only exactly fulfilled for the case of perfect resolution. In reality, elastically scattered events can be identified by allowing $\Delta\varphi$, $\Delta( \tan\vartheta_{1}\tan\vartheta_{2} )$ and $\Delta p$ to deviate from zero only by a small amount that accounts for the resolution of the HADES detector.

\section{Beam and Track Momentum Corrections}
\label{sec:beamtargetconditions}
Prior to the physics analysis, the data were corrected to account for imperfections in the setup, alignment and beam conditions. The detector calibration and alignment were performed with HH data using the methods described in Ref.~\cite{PECHENOVA201540}, followed by a fine-tuning as described in this section. First, a selection of HH events was performed (\cref{subsec:preselection}). These data were used to correct a small offset of the mean beam-target overlap from the nominal transverse location $(x,y) = (\SI{0}{\mm}, \SI{0}{\mm})$ (\cref{subsec:vertexlocation}). Second, a slight misalignment of the beam axis with the $z$-axis was determined and corrected (see \cref{subsec:beamtilt}). Third, a correction was applied for a systematic bias of the reconstructed particle momentum (\cref{subsec:momentumcorrections}).
Finally,  the kinematical constraints of the two-body final state were used to precisely measure the beam energy (\cref{subsec:beamenergy}).

\subsection{Event selection for beam studies}
\label{subsec:preselection}
Here, events are selected that contain two charged tracks in the main HADES spectrometer (HH). Furthermore, the kinematic relations of a two-body reaction are exploited, {\it i.e.} the selected events fulfill the coplanarity condition $|\Delta\varphi| < \ang{0.2}$ and the polar angles relation $|\Delta(\tan\vartheta_{1}\tan\vartheta_{2})| < 0.02$. These criteria are set to be tight, since clean samples are of high importance to eliminate systematic effects in the extraction of correction factors. The selection criteria for the physics analysis described below are different.

\subsection{Vertex Location}
\label{subsec:vertexlocation}
The location of the primary vertex in the transverse plane was determined using the point of closest approach (POCA) between at least two outgoing scattered tracks, applying Tukey-weights. We found a small offset of the beam-target overlap from the nominal position, that was stored in the form of additional track parameter relative to the beam, and taken into account in the analysis. The resulting offsets across the five beam-time periods of the Feb22 campaign were determined with an uncertainty of \SI{0.001}{\mm}, and are presented in \cref{tab:Offsets-N-Tilts}. For reference, the beam-target overlap region has an FWHM of $\pm 2$ mm.

\begin{table}[!b]
 \caption{Overview of the determined mean values of the vertex location $V_x$, $V_y$, and beam misalignment $\alpha_x$, $\beta_y$, $\tan\vartheta_{1}\tan\vartheta_{2}$ from which the beam kinetic energy $T$ is determined.}
 \label{tab:Offsets-N-Tilts}
 \begin{tabularx}{1.0\linewidth}{
  Y
  S[table-format=+1.3]
   @{\hspace{7pt}}
  S[table-format=+1.3]
   @{\hspace{7pt}}
  S[table-format=+1.3]
   @{\hspace{7pt}}
  S[table-format=1.3]
   @{\hspace{7pt}}
  S[table-format=1.3]
   @{\hspace{7pt}}
  S[table-format=1.3]
 }
  \toprule
  Beam   &
  {$\expval{V_x}$} &
  {$\expval{V_y}$} &
  {$\expval{\alpha_x}$} &
  {$\expval{\beta_y}$} &
  {$\expval{\tan\vartheta_{1}\tan\vartheta_{2}}$} &
  {$T$} \\
  period &
  {[\unit{\mm}]} &
  {[\unit{\mm}]} &
  {[\unit{\degree}]} &
  {[\unit{\degree}]} &
  &
  {[\unit{\gev}]} \\
  \midrule
  I      &  0.111  &  -2.950  &  0.085  &  0.074  &  0.293 & 4.529 \\
  IIA    &  2.884  &  -3.369  &  0.053  &  0.049  &  0.293 & 4.529 \\
  IIB    &  2.730  &  -2.252  &  0.024  &  0.051  &  0.293 & 4.529 \\
  III    &  1.891  &  -1.347  &  0.044  &  0.063  &  0.293 & 4.529 \\
  IV     &  2.558  &   0.092  & -0.016  &  0.087  &  0.540 & 1.602 \\
   \bottomrule
 \end{tabularx}
\end{table}

\subsection{Beam Angular Misalignment}
\label{subsec:beamtilt}
In addition to the aforementioned offset, the deviation of the mean proton beam direction from the $z$-axis has been evaluated. The \textit{beam misalignment} is parameterized by the Cartesian angles $\alpha_x$, $\beta_y$ defined as
\[
  \tan \alpha_x = \dv*{y}{z},\quad\tan \beta_y = \dv*{x}{z}.
\]
The mean deviations $\expval{\alpha_x}$ and $\expval{\beta_y}$ are obtained by scanning over $\alpha_x$ and $\beta_y$ in steps of \ang{0.002} and extracting the number of events fulfilling the preselection criteria. The values of $\alpha_x$ and $\beta_y$ yielding the highest event rates are taken as the mean deviation of the beam from the $z$-axis. Individual values of $\expval{\alpha_x}$ and $\expval{\beta_y}$ were determined for the five beam periods of the Feb22 campaign with a precision of \ang{0.001} and are listed in \cref{tab:Offsets-N-Tilts}.

\subsection{Track Momentum Corrections}
\label{subsec:momentumcorrections}

\begin{figure}[!b]
 \includegraphics[width=\linewidth]{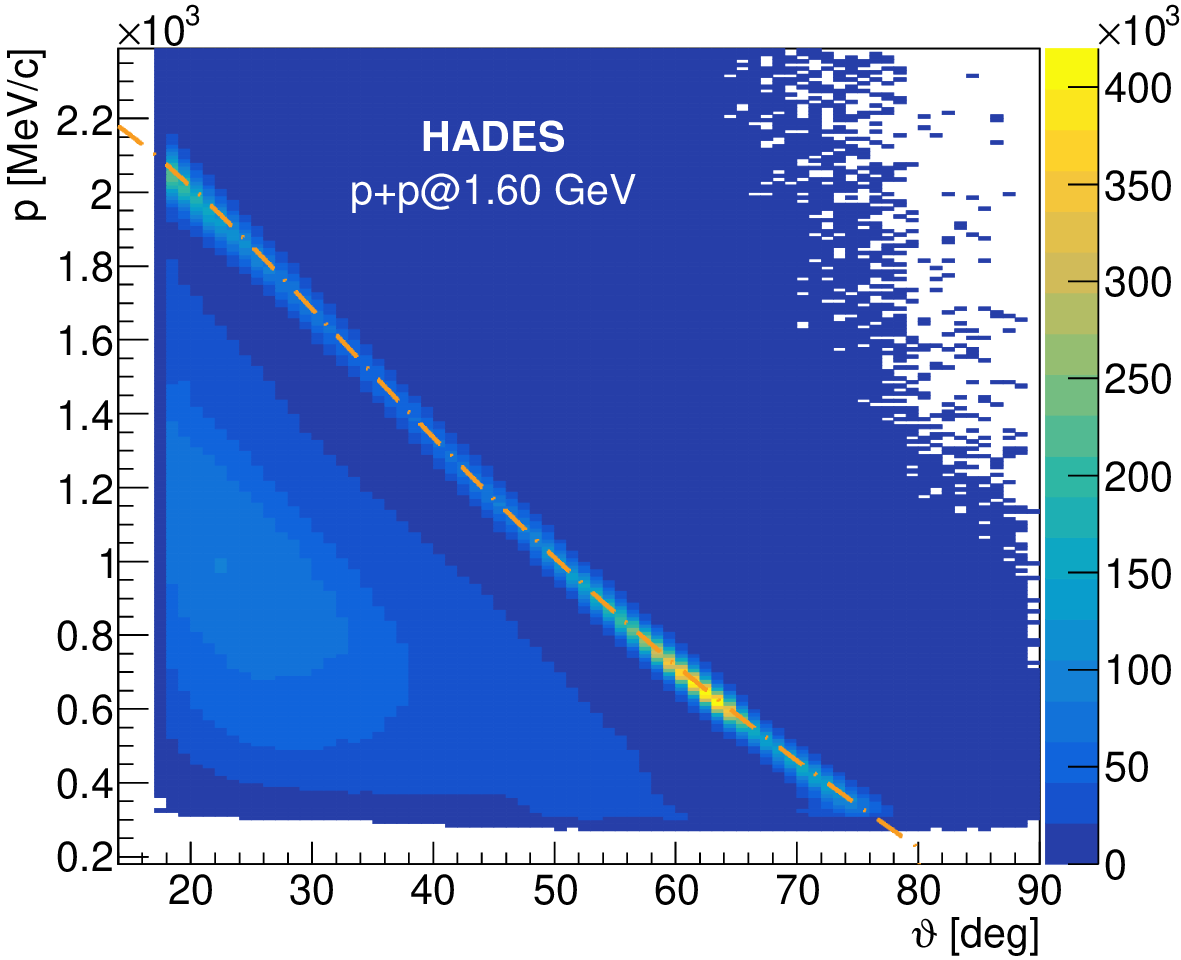}
 \caption{Correlation of momentum versus polar angle for tracks after energy-loss corrections. The orange dot-dashed line indicates the expectation for elastically scattered events, given by \cref{eq:el_poftheta}. }
 \label{fig:PvsTheta}
\end{figure}

Systematic effects arise from the momentum reconstruction due to magnetic field distortions with respect to the map calculated for the HADES toroid magnet. These were investigated by comparing the reconstructed momentum $p_\mathrm{reco}$, determined from a Runge-Kutta fit~\cite{Agakichiev_2009_HADES}, with the expected momentum of an elastically scattered particle $p_\mathrm{theo}$. The latter was calculated from the measured polar angle $\vartheta$ using \cref{eq:el_poftheta}. In a first step, the mean energy loss for the particle in the target and detector material was corrected. Next, the relative deviation $\Delta p/p_\text{theo} = (p_\text{reco} - p_\text{theo})/p_\text{theo}$ was evaluated for elastically scattered protons as a function of $\vartheta$ and the azimuthal angle $\varphi$. The behavior of $\Delta p/p_\text{theo}$ with respect to $\vartheta$ was compared for each sector of the HADES spectrometer, at both beam energies. From this, a correction matrix for the relative momentum was generated. Using interpolation, this matrix was used to determine a momentum correction for each track. The corrections are mostly below \SI{1}{\percent} and always below \SI{4}{\percent}. The standard deviation of $\Delta p/p_\text{theo}$ was determined to be 0.0022 and 0.0058 for the data taken at $T = \SI{1.60}{\gev}$ and \SI{4.53}{\gev}, respectively. The correlation of the corrected momentum versus $\vartheta$ is shown in \cref{fig:PvsTheta} for all tracks recorded during one day, without the preselection. The data exhibit a band from the elastically scattered protons that closely follows \cref{eq:el_poftheta}, indicated by the orange dot-dashed line. This is also the case for angles above \ang{65}, where the energy loss is significant. In addition to the elastically scattered protons, one can discern a broad distribution of mostly pions and protons produced in inelastic collisions.

\begin{figure*}[!ht]
\centering
 \includegraphics[width=\linewidth]{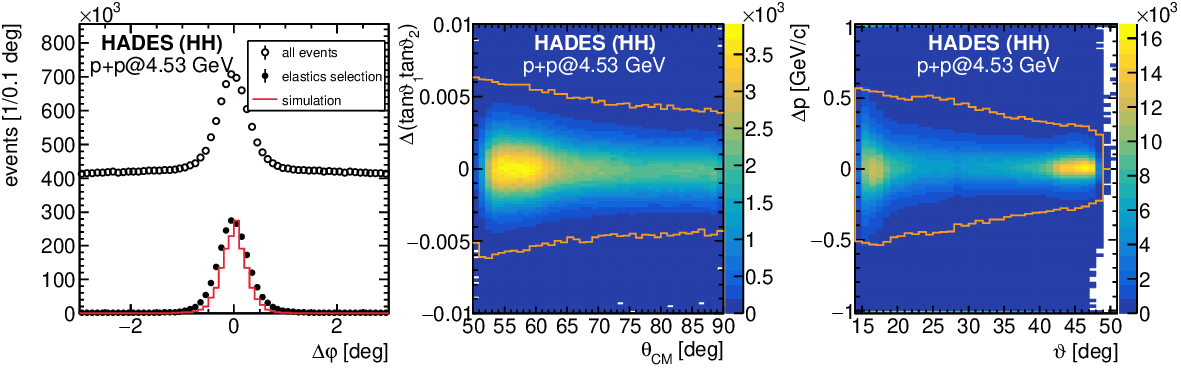}

 \includegraphics[width=\linewidth]{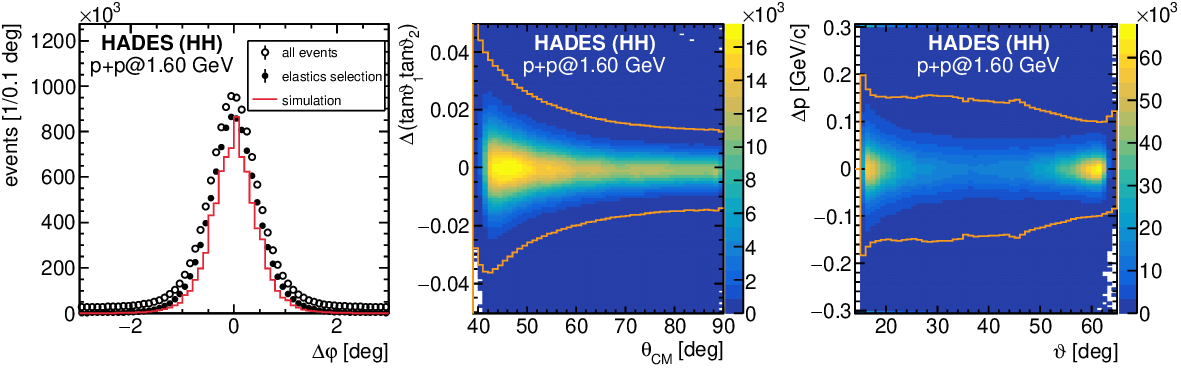}

 \caption{HH selection for $T = \SI{4.53}{\gev}$ (top) and $T = \SI{1.60}{\gev}$ (bottom). Left: Coplanarity of two tracks without (empty circles) and with (filled circles) the $\Delta(\tan\vartheta_{1}\tan\vartheta_{2})$ and $\Delta p$ criteria applied. Center: Selection window for the $\Delta(\tan\vartheta_{1}\tan\vartheta_{2})$ criterion with the coplanarity and $\Delta p$ selection applied. Right: Selection window for $\Delta p$ with the coplanarity and $\Delta(\tan\vartheta_{1}\tan\vartheta_{2})$ criteria applied.}
 \label{fig:hh_selection}
\end{figure*}

\subsection{Beam Energy}
\label{subsec:beamenergy}
After correcting for the actual location of the beam-target overlap and the direction of the beam, \cref{eq:tanthetaproduct} was exploited to investigate the magnitude of the beam energy. The nominal beam energies were $T_\text{\rm{nom}} = \SI{1.58}{\gev}$ and \SI{4.5}{\gev}, but slight variations can occur due to uncertainty of the exact path of the beam through the synchrotron. These manifest in measured values of $\tan\vartheta_{1}\tan\vartheta_{2}$ deviating from the values predicted using the nominal beam energies. Conversely, the beam energy can be extracted from the measured $\tan\vartheta_{1}\tan\vartheta_{2}$ distribution. Thus, the mean value of $\tan\vartheta_{1}\tan\vartheta_{2}$ and its uncertainty were measured for each sector pair. The results were used to determine the value of $T$ and its uncertainty for each HADES beamtime period. The results are summarized in \cref{tab:Offsets-N-Tilts}. These updated beam energies for the full campaign were found to be \SI{1.6018 \pm 0.0006}{\gev} and \SI{4.529 \pm 0.003}{\gev}, in the following referred to as $T = \SI{1.60}{\gev}$ and $T = \SI{4.53}{\gev}$, respectively. The uncertainties have been determined by comparing the three sector pairs. It is not unusual that the requested, nominal beam energies differ from the actual, measured ones, see \textit{e.g.} a measurement at the COSY storage ring \cite{Goslawski:2009vf}. The somewhat larger discrepancies found in this work, can be understood by the SIS18 being designed for heavy-ions fulfilling $Z/A \approx 2$.

\section{Elastic Scattering Selection}
\label{sec:eventselection}

Elastic \prot\prot\decays\prot\prot events, used for luminosity and cross-section measurements, were selected by exploiting the two-body kinematics discussed in \cref{sec:kinrel}. The first two relations (\cref{eq:coplanarity,eq:tanthetaproduct}), require two measured protons, \textit{i.e.} either the HH or the HF category. The third relation (\cref{eq:deltap}) applies to all three data categories, hence also HS. In the following, the event selection for the three cases is described.

\subsection{HH Selection}
\label{subsec:HADESHADES}

\begin{table*}
 \centering
 \caption{Effect of the event selection criteria of the HH analysis, quantified by absolute yields of events surviving each selection criterion, from experimental (column 2 and 5) and simulated (column 3 and 6) data . The acceptance-normalized efficiencies are given in column 4 and 7.}
 \label{tab:el_cut_flow_hh}
 \begin{tabularx}{1.0\linewidth}{
  X
  S[table-format=1.2e1]
   @{\hspace{2em}}
  S[table-format=1.2e1]
   @{\hspace{2em}}
  S[table-format=3.1]
   @{\hspace{5em}}
  S[table-format=1.2e1]
   @{\hspace{2em}}
  S[table-format=1.2e1]
   @{\hspace{2em}}
  S[table-format=3.1]
 }
  \toprule
   Selection &
   {Feb22 Data \SI{4.53}{\gev}} & {Sim \SI{4.53}{\gev}} & {Eff.\,$\%$)} &
   {Feb22 Data \SI{1.60}{\gev}} & {Sim \SI{1.60}{\gev}} & {Eff.\,$\%$}\\
  \midrule
   2 tracks                                  & 6.98e08 & 5.30e06 & 100.0 & 4.92e07 & 5.48e06 & 100.0 \\
   coplanarity                               & 1.06e07 & 1.08e06 &  20.4 & 1.18e07 & 4.28e06 &  78.1\\
   $\Delta(\tan\vartheta_1 \tan\vartheta_2)$ & 2.13e06 & 1.00e06 &  18.9 & 1.0e07  & 4.04e06 &  73.7\\
   $\Delta p$                                & 1.82e06 & 0.97e06 &  18.3 & 8.95e06 & 3.89e06 &  71.0\\
  \bottomrule
 \end{tabularx}
\end{table*}

An overall event CM angle  $\theta_{\mathrm{CM}}$ is defined for the HH events using the CM angles of both measured protons $\theta_{1,2}$ as $\theta_{\mathrm{CM}} = [\theta_1 + (\ang{180} - \theta_2)]/2$. The HH event topology covers $\theta_{\mathrm{CM}}$ in the intervals [\ang{50}, \ang{90}] and [\ang{45}, \ang{90}] for the $T = \SI{4.53}{\gev}$ and \SI{1.60}{\gev} data, respectively. The identification of HH events is based on the relations in  \cref{eq:coplanarity,eq:tanthetaproduct,eq:deltap}. To ensure that the protons are coplanar, $\Delta\varphi$ is required to be within a $3\sigma$ window around the expected value of \ang{0} (\cref{eq:coplanarity}), as indicated in the left panels of \cref{fig:hh_selection}. The selection window was chosen in such a way that the efficiency-corrected signal yield does not depend on the exact value of $\Delta\varphi$, hence where data and simulations are in good agreement.

The $\tan\vartheta_{1}\tan\vartheta_{2}$ resolution was found to have a clear dependence on $\theta_{\mathrm{CM}} = [\theta_1 + (\ang{180} - \theta_2)]/2$. Therefore, the corresponding selection windows depend on $\theta_{\mathrm{CM}}$: the product $\tan\vartheta_{1}\tan\vartheta_{2}$ was required to differ by no more than $3.6 \sigma$ from the mean of the distribution, for all values of $\theta_{\mathrm{CM}}$. The $\sigma$ and mean values were obtained from a fit of a Gaussian function in bins of $\theta_{\mathrm{CM}}$, and the size of the selection windows was chosen to minimize the dependence of the efficiency-corrected signal yield on the window size. In the middle panels of \cref{fig:hh_selection}, the $\Delta(\tan\vartheta_{1}\tan\vartheta_{2})$ selection windows are shown for both beam energies. 
Similarly, the resolution of the momentum difference $\Delta p$ depends on the proton scattering angles $\vartheta$ in the laboratory frame. Hence, $\Delta p$ was required to be within $3.6 \sigma$ from the mean value of the distribution, as shown in the right panels of \cref{fig:hh_selection}.

The aforementioned selection criteria were found to reduce the background so efficiently that no particle identification was required for HH events. The yield of elastic \prot\prot\decays\prot\prot events was determined in bins of $t$, each with a size of \\\SI{0.15}{{\sgevc}}, by integrating the coplanarity peak (left panel of \cref{fig:hh_selection}) across the 3$\sigma$ window. The resulting yields, after applying the selection criteria consecutively, are summarized in \cref{tab:el_cut_flow_hh}. The effect of the first two selection criteria is much larger for experimental data compared to simulated data. This is because experimental data contain inelastic events, whereas the simulated data only contain elastically scattered events, including subsequent interactions.

\subsection{Proton PID for HS and HF Selection}
\label{subsec:protonPID}

\begin{figure}[!t]
  \centering
  \includegraphics[width=1.0\linewidth]{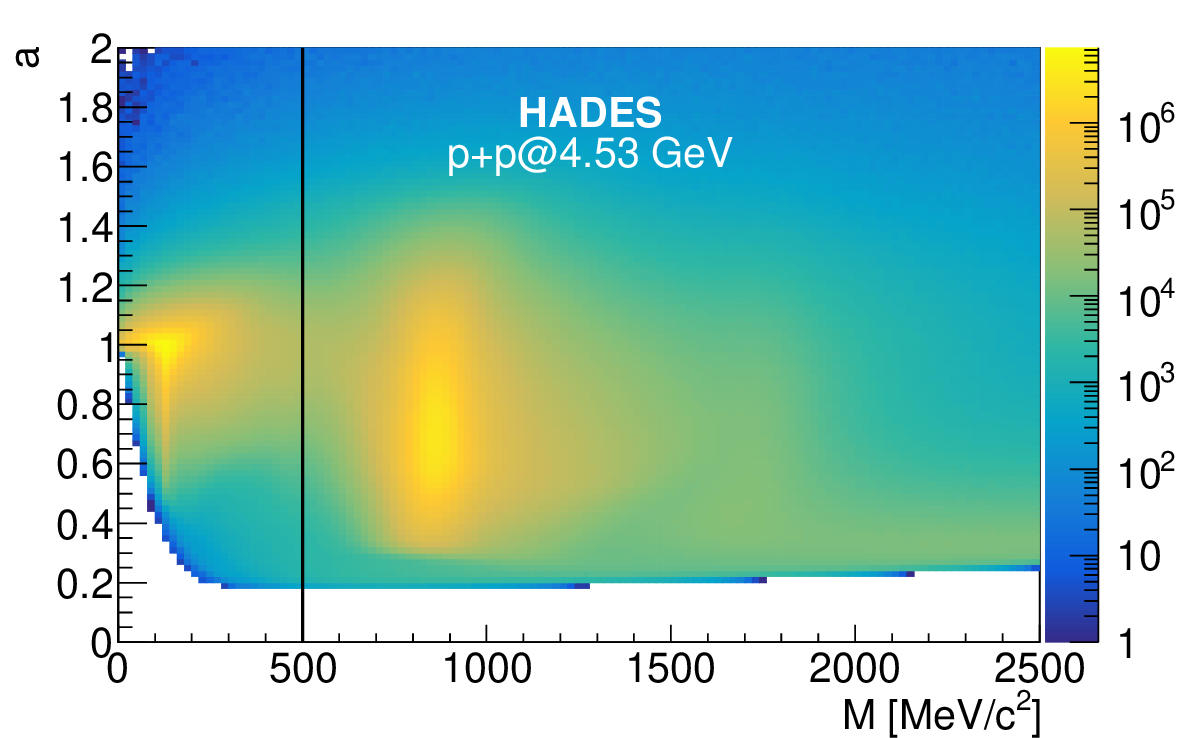}
  \caption{The kinematic variable $a$ vs. the reconstructed mass $M$ for tracks produced at $T = \SI{4.53}{\gev}$. The black line indicates the PID selection, which rejects tracks with $M< \SI{500}{\mevsc}$.}
  \label{fig:pid_mass}
\end{figure}

In contrast to the HH events, the kinematic conditions alone are not sufficient to select pure elastic scattering events in the HS and HF cases. A particle identification (PID) criterion was therefore applied to ensure that the selected particles were, to a large extent, protons. Protons were selected based on the reconstructed mass of positively charged particle tracks. The mass was calculated from the momentum $p$, given by the curvature of the track in the magnetic field, and the velocity $\beta c$, given by the length of the trajectory and the measured time-of-flight:
\[
  M = \frac{p}{c} \sqrt{\frac{1}{\beta^2}-1}.
\]
Tracks fulfilling $M>\SI{500}{\mevsc}$ were selected, primarily to suppress pions. This PID criterion was applied before the energy-loss correction was performed, since the energy loss depends on the particle type. The kinematic variable $a = \sqrt{1+b^2p^2-(1-\beta^2)^2}$ is suitable to investigate the dependence of the reconstructed mass on the momentum and the velocity of the particle. The coefficient $b^2=\SI[per-mode=power]{4e-7}{\per\mev\squared\clight\squared}$ is introduced to compensate for the dimensional difference of these variables.
\begin{figure}[!t]
  \centering
  \includegraphics[width=1.0\linewidth]{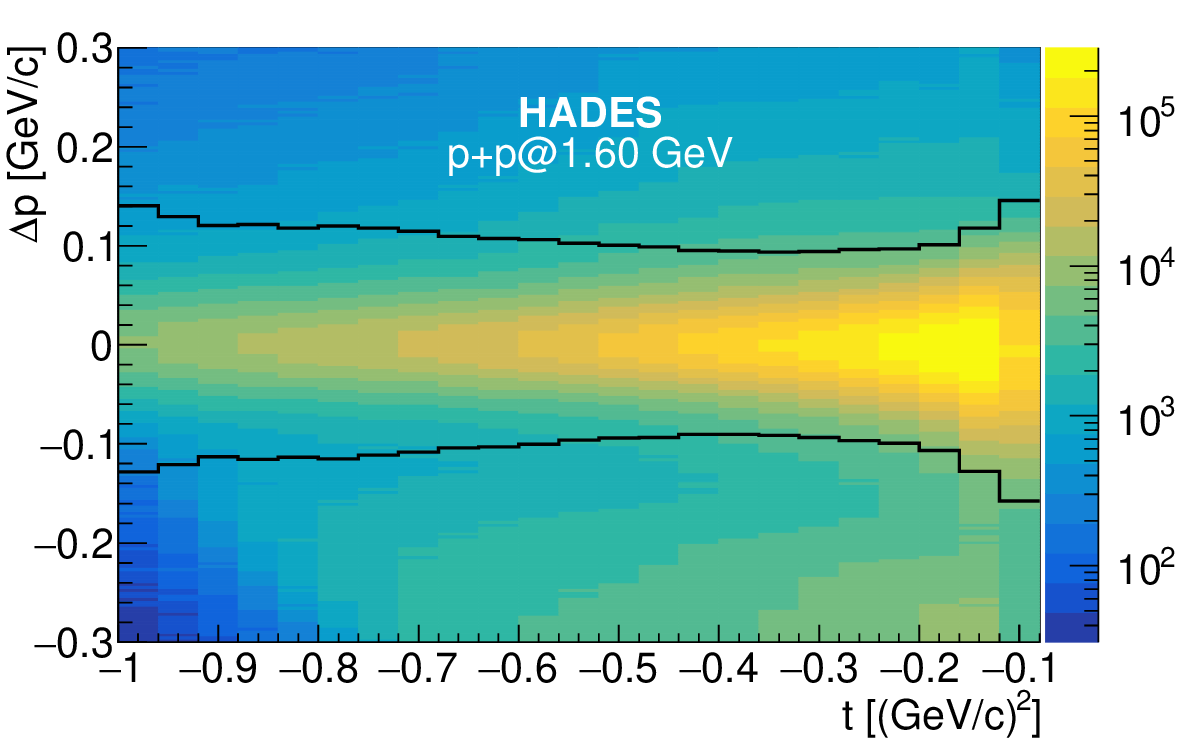}
  \caption{$\Delta p$ vs. $t$ for HS events in the $T = \SI{1.60}{\gev}$ data set. The $t$-dependent selection window is shown in black.}
  \label{fig:deltap_h}
\end{figure}
\begin{figure}[!t]
  \centering
  \includegraphics[width=1.0\linewidth]{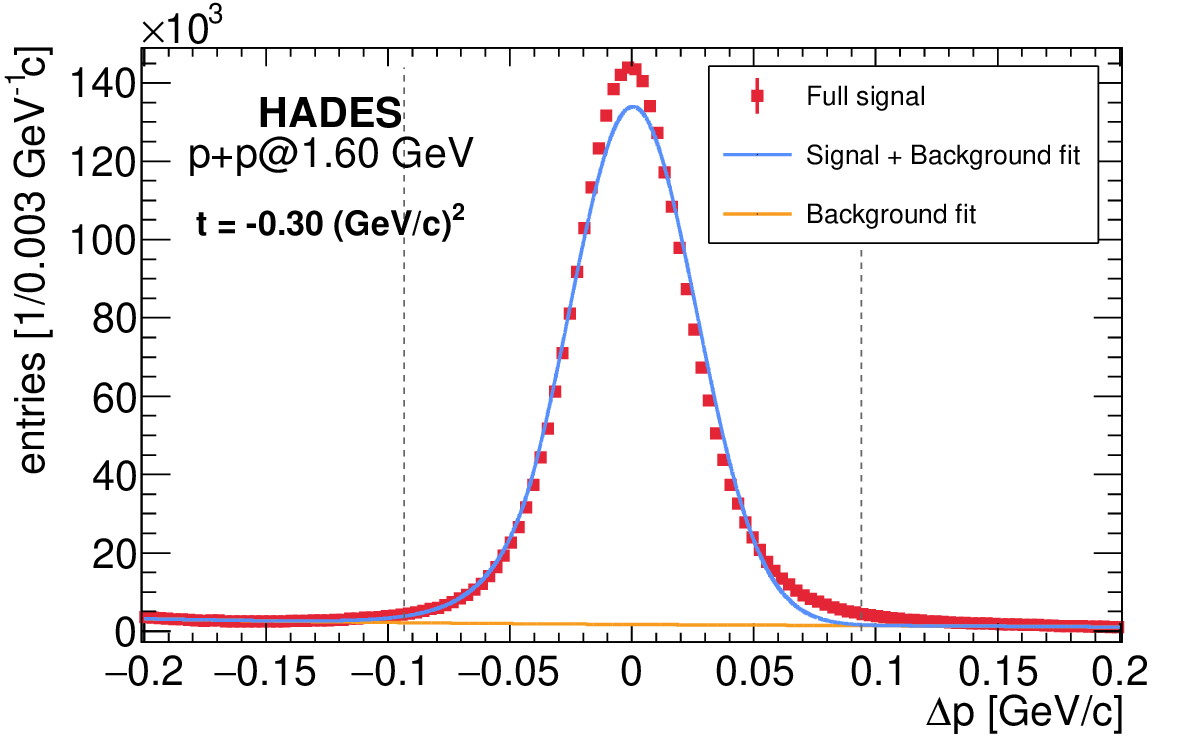}
  \caption{$\Delta p$ projection of \cref{fig:deltap_h} in the bin of width $\SI{0.04}{{\sgevc}}$ at $t=\SI{-0.3}{{\sgevc}}$. The blue line indicates the signal + background fit. Background was fitted outside of the peak area. The signal fit defines the $3.6\sigma$ selection window, indicated with gray dashed lines.}
  \label{fig:deltap_h_proj}
\end{figure}
\begin{figure*}
  \centering
  \includegraphics[width=\linewidth]{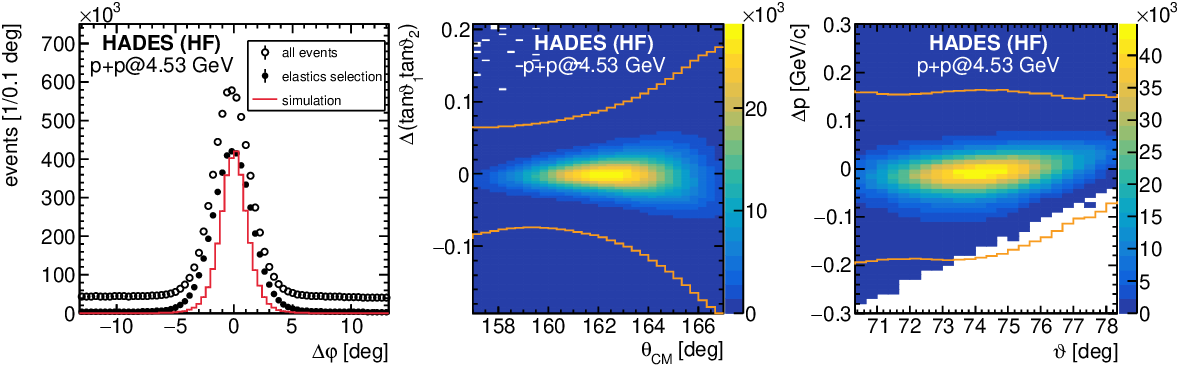}
  \caption{HF selection for \SI{4.53}{\gev} run. Left: Coplanarity of two tracks, with (filled circles) and without (empty circles) the HF selection. Center: Selection window for the $\Delta(\tan\vartheta_{1}\tan\vartheta_{2})$ variable, with the $\Delta p$ condition applied. Right: Selection window for the $\Delta p$ variable, with the $\Delta(\tan\vartheta_{1}\tan\vartheta_{2})$ condition applied.}
  \label{fig:hf_selection}
\end{figure*}
For tracks with $|\Delta p|<\SI{0.3}{\gevc}$, the proton peak is clearly visible in the $a$ versus $M$ distribution shown in \cref{fig:pid_mass}.
 This chosen selection separates protons from pions for all $a$. All FD tracks were assumed to be protons. This turned out to be a good approximation since the much lighter pions are typically emitted at polar angles not covered by the FD, and kaons are so rare that they can be neglected.

\subsection{HS Selection}
\label{subsec:HADESsingles}

\begin{table}[!b]
  \centering
  \caption{Effect of the HS selection criteria at $T = \SI{1.60}{\gev}$, quantified by absolute yields of events surviving each selection criterion, from experimental and simulated data (columns 2 and 3). The acceptance-normalized efficiencies are also given (column 4).}
  \label{tab:el_cut_flow_h}
  \begin{tabularx}{1.0\linewidth}{
   X
   S[table-format=1.2e1]
    @{\hspace{2em}}
   S[table-format=1.2e1]
    @{\hspace{2em}}
   S[table-format=3.1]
   }
   \toprule
    Selection & {Data} & {Sim} & {Eff.\,($\%$)}  \\
   \midrule
    1 positive track        & 7.84e07   & 1.95e07   & 100.0      \\
    $M > \SI{500}{\mev}$    & 6.06e07   & 1.91e07   & 97.9  \\
    $\Delta p < 3.6\sigma$  & 4.50e07   & 1.84e07   & 94.4  \\
    Background subtraction  & 4.29e07   &  1.84e07  &  94.4     \\
   \bottomrule
  \end{tabularx}
\end{table}

In this case, only one proton was detected and as a consequence, neither $\Delta \varphi$ nor $\Delta(\tan\vartheta_{1}\tan\vartheta_{2})$ can be calculated. Therefore, a criterion on $\Delta p$, defined by \cref{eq:deltap}, was applied in combination with the proton PID, as presented in \cref{subsec:protonPID}. \Cref{fig:deltap_h} shows $\Delta p$ as a function of $t$ for HS events from the $T = \SI{1.60}{\gev}$ data set. A fit was performed to the $\Delta p$ distribution using a Gaussian function with an asymmetric tail on top of a third-order polynomial describing the background. In \cref{fig:deltap_h_proj}, an example projection of the $\Delta p$ distribution is shown for one $t$ bin. The signal yield is extracted by subtracting the background and integrating the $\Delta p$ distribution within a range defined by $\pm 3.6\sigma$ around the peak position, in $t$ bins of size \SI{0.04}{{\sgevc}}. The size of $\pm3.6\sigma$ was chosen to minimize the dependence on the corrected signal yield on the window size. At $T = \SI{1.60}{\gev}$, the HS selection resulted in a clean sample of elastic \prot\prot\decays\prot\prot events. The signal yield for each selection step of the HS analysis is summarized in \cref{tab:el_cut_flow_h}. At $T = \SI{4.53}{\gev}$, the background contribution from inelastic channels turned out to be large, \textit{i.e.} more than $10\,\%$ of the signal yield for $|t|>\SI{0.35}{{\sgevc}}$. Furthermore, the background shape was difficult to parameterize, resulting in a large model dependence for the background subtraction. Therefore, only the HS data collected at $T = \SI{1.60}{\gev}$ were used.

\subsection{HF Selection}
\label{subsec:HADESForward}

\begin{table}[!b]
  \centering
  \caption{Effect of the HF selection criteria at $T = \SI{4.53}{\gev}$, quantified by absolute yields of events surviving each selection criterion, from experimental and simulated data.}
  \label{tab:el_cut_flow_hf}
  \begin{tabularx}{1.0\linewidth}{
   X
   S[table-format=1.2e1]
    @{\hspace{2em}}
   S[table-format=1.2e1]
    @{\hspace{2em}}
   S[table-format=3.1]
  }
   \toprule
    Selection & {Data} & {Sim} & {Eff.\,($\%$)}  \\
   \midrule
    1 pos. track in HADES   & 2.55e09   & 1.51e08    & 100.0    \\
    $M>\SI{500}{\mev}$      & 1.72e09   & 1.50e08    & 99.3     \\
    FD coincidence          & 9.31e08   & 8.88e07    & 58.8     \\
    $\Delta(\tan \vartheta_1 \tan \vartheta_2)$
                            & 2.39e08   & 8.40e07    & 55.6     \\
    $\Delta p$              & 2.04e08   & 8.35e07    & 55.3     \\
   \bottomrule
  \end{tabularx}
\end{table}

The HF selection requires one of the scattered protons to enter the main HADES acceptance, while the other is detected in the FD system. For a track to be reconstructed in the FD, at least one hit in each double layer is required. This implies a minimum of four hits in each STS (\cref{sec:experimentalsetup}). The proton candidate in the main HADES spectrometer was identified using the PID criterion in \cref{subsec:protonPID}, and then combined with the proton detected in the FD system. 

Next, events were selected within a 5.5$\sigma$ window around the mean of the $\Delta( \tan \vartheta_{1}\tan\vartheta_{2})$ distribution, in bins of $\theta_{\mathrm{CM}}$ (see \cref{fig:hf_selection}, middle panel).  Furthermore, $\Delta p$ was required to be within a $4.5\sigma$ window at a given $\vartheta_i$ distribution (see \cref{fig:hf_selection}, right panel). The window sizes were obtained by maximizing the figure-of-merit given by $S/\sqrt{S+B}$, where $S$ is the signal yield and $B$ the background yield. It was also checked that in this region, the efficiency-corrected signal yield does not depend on the window size. In events where pile-ups cause multiple HF pair candidates, the best elastic pair candidate was identified using the quadratic sum of the relative deviations of $\Delta p$, $\Delta( \tan \vartheta_\mathrm{1} \tan\vartheta_\mathrm{2})$ and $\Delta \varphi$:
\begin{equation}
  \label{eq:deltas}
  \delta_{\text{total}} = \sqrt{ \left( \frac{\Delta p}{\delta_{\Delta p}} \right)^2 + \left( \frac{\Delta \varphi}{\delta_{\Delta \varphi}} \right)^2 + \left( \frac{\Delta( \tan \vartheta_\mathrm{1} \tan\vartheta_\mathrm{2}) }{ \delta_{\tan \vartheta_\mathrm{1} \tan\vartheta_\mathrm{2}} } \right)^2 }.
\end{equation}
The pair with the smallest value of $\delta_\mathrm{total}$ value was retained for further analysis.

The HF signal yield was determined in $t$-bins of size of \SI{0.01}{{\sgevc}}, by integrating the $\Delta\varphi$ distribution between \ang{170} and \ang{190}. The analysis was restricted to the $t$-region $\SI{-0.2}{{\sgevc}} \leq t \leq \SI{-0.1}{{\sgevc}}$, where the background is negligible and hence no background subtraction was needed. In \Cref{tab:el_cut_flow_hf}, the effect of the HF selection criteria is summarized for data and simulation. The efficiency drop which occurs when introducing the FD coincidence is due to the limited acceptance of the FD. At $t \leq \SI{-0.2}{{\sgevc}}$, the background increased significantly so that subtracting it led to a model-dependent yield of signal events. Therefore, this $t$-region is excluded from further analysis.

\section{Experimental effects}
\label{sec:acceptanceefficiency}

\subsection{Efficiency corrections}
\label{subsec:corrections}
The efficiency as a function of the momentum transfer $t$ has been studied using MC simulations within the $t$-range relevant for this measurement. For this, the same analysis procedure was used as for experimental data. The combined acceptance and reconstruction efficiency $\varepsilon$ is defined as
\begin{equation}
  \varepsilon = N_\mathrm{el,reco}/N_\mathrm{el,MC},
  \label{eq:eff}
\end{equation}
where $N_\mathrm{el,reco}$ is the number of successfully reconstructed elastic events from the simulations and $N_\mathrm{el,MC}$ the total number of generated MC events. This efficiency was calculated in each $t$-bin and was found to be between $\approx\SIrange{10}{80}{\percent}$. From the efficiency in each bin, correction factors were calculated and applied to the experimental data. This $t$-dependent efficiency correction minimizes the model dependence of the correction arising from differences in the simulated and true shape of the differential cross section.

\subsubsection{HF correction factor}\label{subsec:hf_corrections}

To validate and improve the efficiency correction for the newly installed FD, the HS and HF yields have been compared for both experimental data and simulations in the region {\it i.e.} $\SI{-0.2}{{\sgevc}} < t < \SI{-0.1}{{\sgevc}}$, where their acceptances overlap. Ideally, with a perfect MC model, the ratio of the exclusive HF and the inclusive HS yields should be the same in simulations and experimental data. The ratio $\eta$ is defined as
\[
\label{eq:eta_factor}
    \eta = \frac{(N_\mathrm{HF} / N_\mathrm{HS})_\mathrm{sim}}{(N_\mathrm{HF} / N_\mathrm{HS})_\mathrm{data}},
\]
to quantify imperfections in the MC description. It was found that in this region, the value of $\eta$ is nearly constant and close to one. To account for the small deviations (within 5\,$\%$), the efficiency values were corrected by the $\eta$ factor in each $t$-bin.

\subsubsection{Sector Exclusion and Scaling}
The reconstruction efficiency $\varepsilon$ and the correction factor $\eta$ were found to exhibit a consistent $t$-dependence in each HADES sector. However, for one sector, the $t$-dependence of the corrected HF signal yield was found to be significantly different. Due to the cylindrical symmetry of the scattering process, any physical effect must be independent of the azimuthal angle $\varphi$. Therefore, the origin of the observed inconsistency must be related to the experimental setup. The effect was isolated to the outer edge of the HADES acceptance for cases where one proton enters the FD. It did not appear in the HH nor the HS analyses, in which cases the acceptance does not cover the outer edge of the HADES sectors. The inconsistent sector was excluded from the HF final analysis and the total yield was subsequently scaled by a factor of $6/5$, assuming azimuthal asymmetry. This assumption was supported by systematic comparisons of the other sectors, and the deviations are included in the systematic error. 

\subsubsection{Target Window Contribution}
\label{subsec:target_window}

\begin{figure}[!t]
  \centering
  \includegraphics[width=1.0\linewidth]{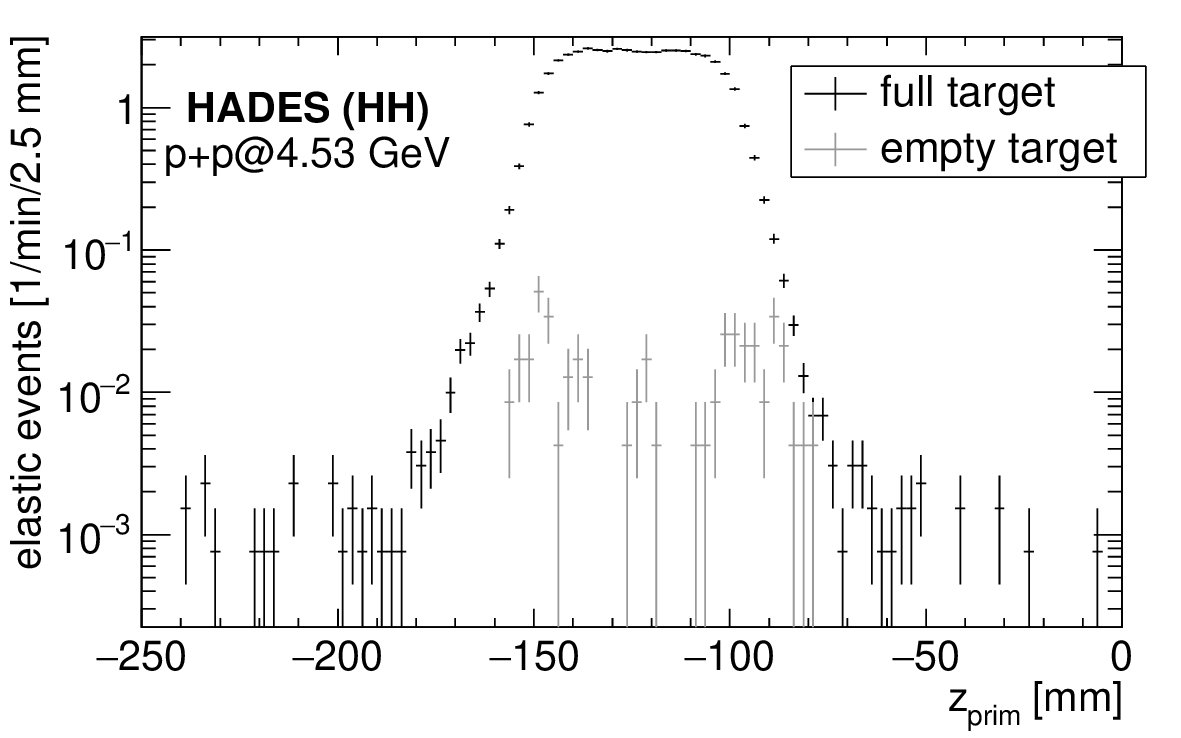}

  \includegraphics[width=1.0\linewidth]{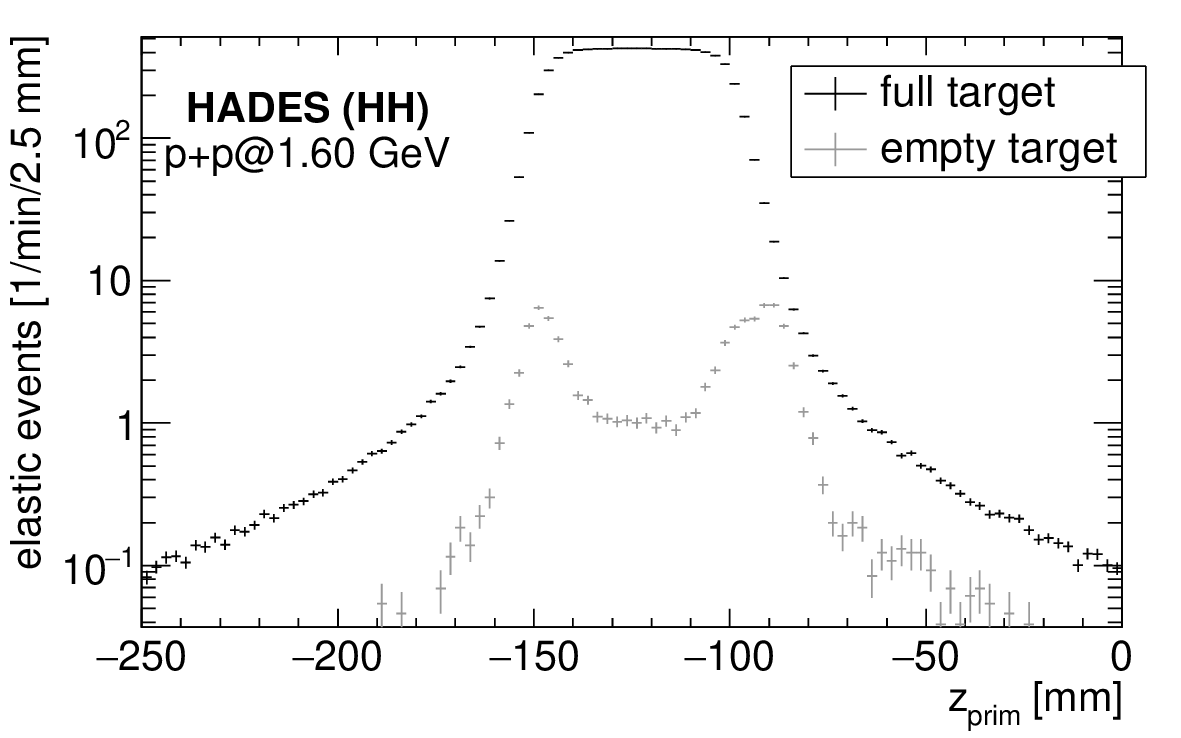}

  \caption{Vertex $z$-position of HH elastic events for full and empty target data, normalized by recorded time.}
  \label{fig:vertex_hh}
\end{figure}

\begin{figure}[!t]
  \centering
  \includegraphics[width=1.0\linewidth]{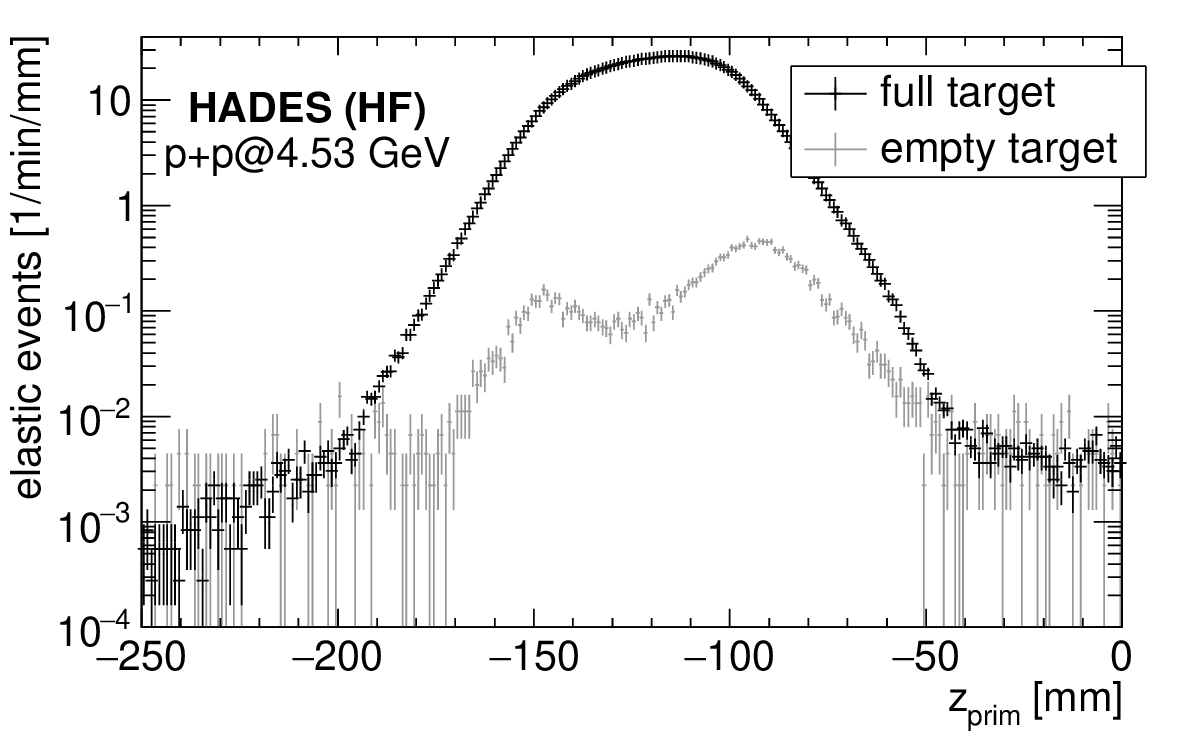}
  \caption{Vertex $z$-position of HF elastic events for full and empty target data, normalized by recorded time.}
  \label{fig:vertex_hf}
\end{figure}

When the beam traverses the entrance and exit windows of the liquid hydrogen target container, it has a \SI{0.05}{\percent} probability to interact with the Mylar foil. The effects of these interactions on the elastic scattering analysis were investigated by studying reference data collected with an empty target. \Cref{fig:vertex_hh} shows the $z$-coordinate distribution of the interaction vertex, $z_\mathrm{prim}$, reconstructed with the POCA method for two protons passing the HH selection for two cases: the full target and the empty target. The entrance and exit windows are clearly discerned in the empty target run. The larger background contribution from the downstream end of the target results from the thermal insulation window mentioned as the “additional foil” in Section 2.1. It was concluded that the HH selection successfully rejects nearly all of the background from non-elastic and non-pp scattering. The total contribution of target window interactions to elastic scattering was found to be \SI{0.8 \pm 0.1}{\percent} at $T = \SI{4.53}{\gev}$ and \SI{1.07 \pm 0.01}{\percent} at $T = \SI{1.60}{\gev}$. The contribution of the empty target was also determined for the HF case. \Cref{fig:vertex_hf} shows the vertex $z$ distribution, reconstructed by determining the POCA between the track reconstructed in HADES after the HF selection and the beamline. Similar to the HH case, the target windows are visible. The contribution of the empty target in the HF selection was determined to be \SI{1 \pm 0.1}{\percent} at $T = \SI{4.53}{\gev}$. 

\section{Systematic Uncertainties}
\label{sec:systematics}

The amount of data available for this measurement is sufficiently high that systematic uncertainties become the dominant contribution to the overall uncertainty of the measured recorded time-integrated luminosities (\cref{sec:luminosity}) and the $t$-dependent cross-section (\cref{sec:crosssectionmeasurements}).

Various sources of systematic uncertainty have been considered in this analysis. Systematic effects arise when correcting the data for detector acceptance and reconstruction efficiency. The associated uncertainties are specific to this experiment. In addition, previous measurements of \prot\prot\decays\prot\prot cross sections have been used as reference when normalizing our data and extracting the luminosity. The uncertainties of previous measurements propagate to our normalization factors and are essentially beyond the control of this analysis. In this section, investigations of systematic effects and uncertainties for each of the three event categories HH, HS and HF, are presented.

\subsection{Systematic Uncertainties in the HH Analysis}
\label{sec:uncertainty_hh}

\textit{Sector Response:}
The efficiency of the six HADES sectors varies with time. To quantify systematic inefficiencies not modeled in the simulations, the HH signal yields are compared for the three different pairs of HADES sector. This source of uncertainty is referred to as \textit{Sectors} in \cref{tab:el_systematics_hh}.

\textit{Selection Criteria:}
Differences in the detector response between experiment and MC simulations manifest in differences in the efficiency-corrected signal yields when varying the selection windows. Therefore, the selection windows for the $\Delta(\tan\vartheta_1\tan\vartheta_2)$ and $\Delta p$ variables are varied (both listed in \cref{tab:el_cut_flow_hh}) in steps of $0.2\sigma$ between $3.2\sigma$ and $4\sigma$ around the peak position. The differences are quoted as systematic uncertainties.

\textit{Coplanarity Fit Range:}
The HH signal yield was evaluated by integrating the fit to the coplanarity peak. The uncertainty was estimated by varying the integration range between $2\sigma$ and $4\sigma$ in steps of $0.2\sigma$, simultaneously in simulation and data. The mean of the differences was calculated and quoted as the systematic uncertainty. 

As can be observed from \cref{tab:el_systematics_hh}, the uncertainties from the sector efficiency and the selection criteria are at the percent level. All listed uncertainties are considered independent of $t$, since the yields were found to vary in a very similar way in all bins when varying the selection criteria.

\textit{Normalization:}
The luminosity at $T = \SI{4.53}{\gev}$ was used for normalizing the HH data to extract the cross sections. This luminosity was determined from the HF sample, where the data points overlap with the measurements from Argonne. The uncertainty of the normalization factor has two main sources: one from the HF analysis, which amounts to \SI{3.2}{\percent}, and the other from the Argonne measurements, which is \SI{4.0}{\percent}. The HF uncertainties are discussed in \cref{sec:uncertainty_hf} and summarized in \cref{tab:el_systematics_hf}. The total uncertainty from the normalization was obtained by adding the two uncorrelated contributions in quadrature and was found to be $5.1\,\%$, hence the largest uncertainty of the cross-section measurements for the HH case.

At $T = \SI{1.60}{\gev}$, the normalization is performed using the SAID fits SM16 and WF16. These have small uncertainties, below $0.06\,\%$, and the two models were found to give very similar results. At this energy, the uncertainty arising from the normalization is dominated by the $t$ range used to normalize the HH signal yields to the differential cross sections from SAID. To quantify this uncertainty, the range has been varied in steps between $t = \brange{-1.42}{-0.38}{{\sgevc}}$ and $t = \brange{-1.30}{-0.50}{{\sgevc}}$. The uncertainty arising from the normalization was found to be $0.8\,\%$.

\textit{Total Uncertainty:}
The contributions to the total systematic uncertainty of the luminosity and cross sections obtained with the HH data are summarized in~\cref{tab:el_systematics_hh}. The total systematic uncertainty was calculated as the quadratic sum of the individual contributions. 

\begin{table}[!b]
 \caption{Systematic uncertainties associated to the HH analysis.}
 \label{tab:el_systematics_hh}
 \centering
 \begin{tabularx}{1.0\linewidth}{
    p{3cm}
    S[detect-weight,mode=text]
    S[detect-weight,mode=text]
  }
  \toprule
   Source & {uncertainty [\si{\percent}]} & {uncertainty [\si{\percent}]} \\
          & {($T = \SI{4.53}{\gev}$)} & {($T = \SI{1.60}{\gev}$)} \\
  \midrule
   Sectors                                  & 1.50   & 0.53 \\
   $\Delta p$ window                        & 0.31   & 0.30 \\
   $\tan\vartheta_1\tan\vartheta_2$ window  & 0.05   & 0.05 \\
   Coplanarity fit range                    & 0.08   & 0.04 \\
  \midrule
   \textbf{Total (selection)}               & \B 1.53 & \B 0.61 \\
  \midrule
   Normalization                            & 5.1    & 0.8 \\
  \midrule
   \textbf{Total}                           & \B 5.3 & \B 1.0 \\
  \bottomrule
 \end{tabularx}
\end{table}

\begin{table}[!hb]
 \caption{Systematic uncertainties associated to the HS analysis.}
 \label{tab:el_systematics_h}
 \centering
 \begin{tabularx}{1.0\linewidth}{
    p{3.5cm}
    S[detect-weight,mode=text]
  }
  \toprule
   Source & {uncertainty [\si{\percent}] ($T = \SI{1.60}{\gev}$)} \\
  \midrule
   Sectors                      & 2.90 \\
   PID                          & 0.33 \\
   $\Delta p$ fit range         & 1.15 \\
  \midrule
   \textbf{Total (selection)}   & \B 3.1 \\
  \midrule
   Normalization                & 1.0 \\
  \midrule
   \textbf{Total}               & \B 3.3 \\
  \bottomrule
\end{tabularx}
\end{table}

\begin{table}[!hb]
 \caption{Systematic uncertainties associated to the HF analysis.}
 \label{tab:el_systematics_hf}
 \centering
 \begin{tabularx}{1.0\linewidth}{
    p{3.5cm}
    S[detect-weight,mode=text]
  }
  \toprule 
   Source & {uncertainty [\si{\percent}] ($T = \SI{4.53}{\gev}$)} \\
  \midrule
   Sectors                                  & 3.18 \\
   $\Delta p$ window                        & 0.001 \\
   $\tan\vartheta_1\tan\vartheta_2$ window  & 0.001 \\
   Coplanarity fit range                    & 0.12 \\
  \midrule
   \textbf{Total (selection)}               & \B 3.18 \\   
     \midrule
   Normalization                            & 4.0 \\
  \midrule
   \textbf{Total}                           & \B 5.11 \\
  \bottomrule
\end{tabularx}
\end{table}

\subsection{Systematic Uncertainties in the HS Analysis}

The uncertainties from inefficiencies in the HADES sectors and from selection criteria are evaluated in the same way as for the HH analysis.

\textit{Sector Response:}
The uncertainty was quantified by comparing the HS signal yield in individual sectors to the average yield. In the HS case, this was found to be the largest contribution to the systematic uncertainty.

\textit{PID:}
The uncertainty from the PID selection was estimated by varying the minimum mass $M$ in 10 steps between \\$\SI{400}{\mevsc}$ and $\SI{600}{\mevsc}$.

\textit{Background subtraction:}
The signal yield was estimated by integrating the $\Delta p$ peak, which was found to be on top of a small background of about $4$--$8\,\%$. The background was parameterized by a polynomial function. The uncertainty in the background estimation was determined by varying the background function and the fit range. This uncertainty was found to be $t$-dependent since the background varies with $t$.

\textit{$\Delta p$ selection:}
The integration window of the $\Delta p$ peak was varied in steps of $0.2\sigma$ between $3.2\sigma$ and $4\sigma$ around the peak position. Differences to the nominal selection window of $3.6\sigma$ are quoted as systematic uncertainty.

\textit{Normalization:}
The $T=\SI{1.60}{\gev}$ data obtained with the HS analysis are normalized using HH data at the same energy, and SAID fits. Therefore, the normalization has a contribution from the uncertainty in the HH signal yield, found to be \SI{0.6}{\percent}. 

\textit{Total Uncertainty:}
The different contributions to the total systematic uncertainty of the HS data are summarized in \cref{tab:el_systematics_h}, and summed in quadrature to obtain the total uncertainty. 

\subsection{Systematic Uncertainties in the HF Analysis} \label{sec:uncertainty_hf}

\textit{Sector response:}
Sector differences were evaluated in the same way as in the HS case, with sector four excluded as motivated in \cref{sec:acceptanceefficiency}.

\textit{Alignment:}
A separate analysis of these data using an improved detector alignment resulted in an increase of the track finding efficiency on the edges of the fiducial volume for the FD. However, in the region included in the luminosity estimation, no significant effect on the determined integrated luminosity was found.

\textit{Selection criteria:}
In the HF analysis, the uncertainties from the $\Delta(\mathbf{\tan\vartheta_1\tan\vartheta_2})$ selection was estimated by varying the selection window from $4\sigma$ to $6.5\sigma$. In a similar way, the uncertainty from the $\Delta p$ window was estimated by varying the window from $3\sigma$ to $5.5\sigma$.

\textit{Coplanarity fit range:}
The HF signal yield was estimated by integrating the fit to the coplanarity peak. The uncertainty was estimated by varying the fit range within data-driven limits based on inspection of the event distributions to ensure good signal quality and sufficient statistical precision. 

The selection windows and the integration range of the coplanarity distribution were found to have a negligible effect on the total systematic error.  All the HADES-specific systematic uncertainties were found to be independent of $t$.

\textit{Normalization:}
The uncertainty in the normalization comes from the limited precision of the Argonne data \cite{PhysRevD.9.1179}, which were used as a reference.

\textit{Total Uncertainty:}
All systematic uncertainties from the HF analysis are summarized in \cref{tab:el_systematics_hf}. The Argonne and HADES data are completely independent, hence the uncertainties are treated as uncorrelated. The total uncertainty is therefore calculated as the quadratic sum of the HADES-specific sources and the Argonne uncertainty.

\section{Time-integrated Recorded Luminosity}\label{sec:luminosity}

The time-integrated luminosity $\mathcal{L}_\mathrm{int}$ is generally extracted from the number of events $N$ from a certain reaction, divided by the cross-section $\sigma$ of that reaction, the trigger scaling factors $f_\mathrm{scal}$, the DAQ livetime $\tau_\mathrm{DAQ}$, the duty factor $\xi$ of the data collection period, the combined efficiency and acceptance $\varepsilon$ defined in \cref{eq:eff}, which includes the trigger efficiency $\epsilon_\mathrm{trig}$:
\begin{equation*}
    \mathcal{L}_\mathrm{int}=\frac{Nf_\mathrm{scal}}{\sigma_{\mathrm{el}}\tau_\mathrm{DAQ}\xi\varepsilon}.
\end{equation*}
In HADES, the main purpose of measuring the luminosity is to extract cross sections of various proton-induced reactions. Given that the same raw data are used for all measurements (luminosity and differential cross sections), the DAQ livetime and the duty factor are the same irrespective of the studied reaction. Hence, the relevant quantity is the \textit{recorded luminosity}, defined as the time-integrated luminosity multiplied by the DAQ livetime and the duty factor. This recorded, time-integrated luminosity is denoted as $\mathcal{L}$ in the following :
\begin{equation*}
  \mathcal{L}=\mathcal{L}_\mathrm{int} \tau_\mathrm{DAQ}\xi.
\end{equation*}

The luminosity determination here uses the yield $N_\mathrm{el}$ of elastically scattered  \prot\prot\decays\prot\prot events obtained in this analysis, the efficiency $\epsilon$ that includes acceptance, reconstruction efficiency and trigger efficiency. The latter was determined using trigger emulators on the simulated data. The \prot\prot\decays\prot\prot differential cross section has been measured by previous experiments at \textit{e.g.} Argonne and COSY, in kinematic regions close to that of HADES. The time-integrated luminosity of the full beamtime at $T = \SI{4.53}{\gev}$ is obtained using the HF selection and the luminosity determination in the $T = \SI{1.60}{\gev}$ case exploits the HH selection. The latter benefits from the low uncertainties of the HH analysis and the reference data (see \cref{sec:uncertainty_hh}).

The recorded, time-integrated luminosity $\mathcal{L}$ is calculated in a given momentum-transfer squared $t$ bin $i$ using
\begin{equation}
 \label{eq:lumi-sum}
 \mathcal{L} = \frac{f_{scal} \bar{\eta}_{w}}{\sigma_\mathrm{el}} \sum_{i} \frac{N_{\mathrm{el},i}}{\varepsilon_{i}},
\end{equation}
where $N_\mathrm{el}$ is the number of reconstructed elastic proton-proton pairs (see \cref{sec:eventselection}), the factor $f_{scal}=64$ is the scaling factor of the minimum bias trigger (see \cref{subsubsec:trigger}), $\varepsilon$ denotes the combined acceptance times reconstruction efficiency (see \cref{sec:acceptanceefficiency}) and $\bar{\eta}_{w}$ is the weighted average of the correction factor accounting for the difference in HF reconstruction efficiency between simulation and data, as defined in \cref{subsec:hf_corrections}. The latter is only applied at $T = \SI{4.53}{\gev}$, where the HF selection is used. Finally, $\sigma_\mathrm{el}$ is the \prot\prot\decays\prot\prot differential cross section obtained from Argonne for the $T = \SI{4.53}{\gev}$ data and from the SAID fits for the $T = \SI{1.60}{\gev}$ data.

\subsection{Time-integrated luminosity at $T = \SI{4.53 }{\gev}$}

\begin{figure}[!t]
 \centering
 \includegraphics[width=\linewidth]{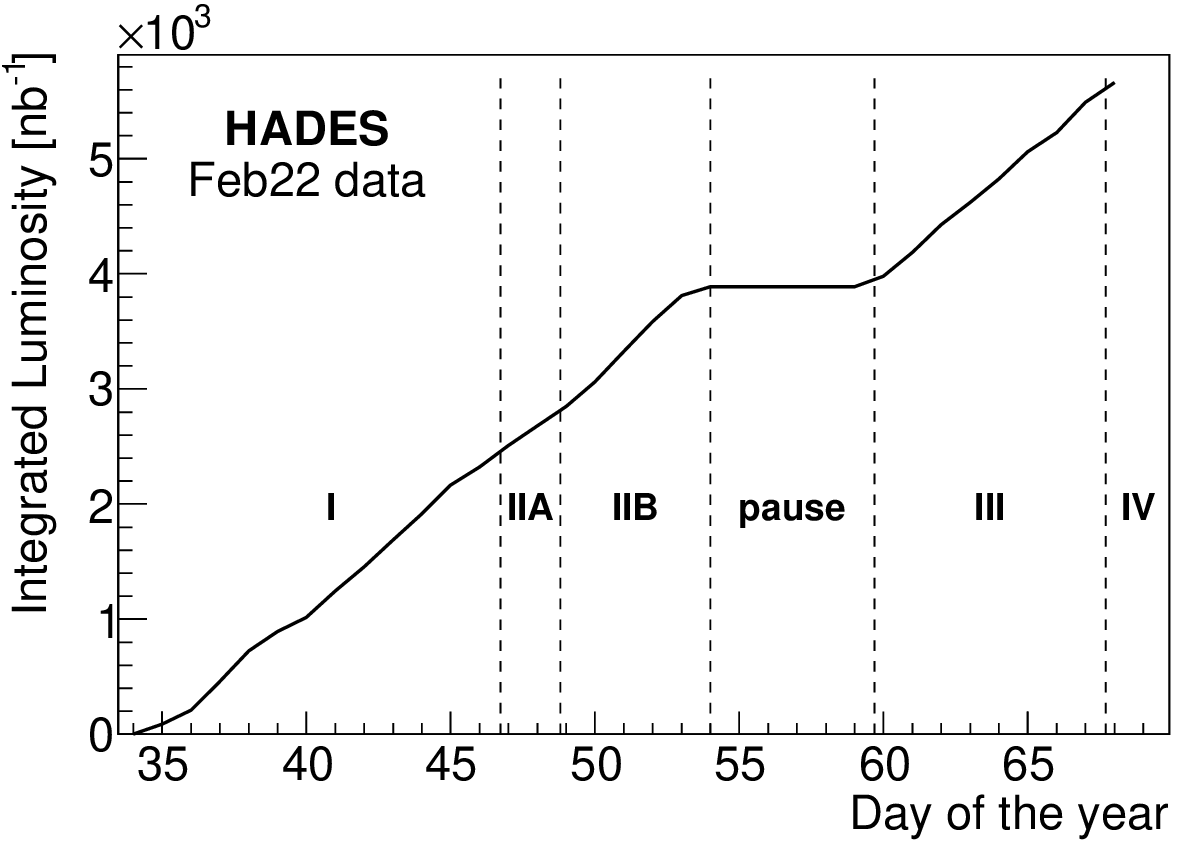}
 \caption{Integrated luminosity for the data taken at $T = \SI{4.53 }{\gev}$  as a function of the day of the year. There was a pause of operation from day 54 to day 59. Dashed lines and labels indicate beam periods as described in \cref{subsec:vertexlocation}.}
 \label{fig:luminosity-high-energy}
\end{figure}

The range $t = \brange{-0.20}{-0.10}{{\sgevc}}$ was chosen for normalization at this energy. Since no previous measurements exist at $T = \SI{4.53}{\gev}$ ($p_\mathrm{beam} = \SI{5.3}{\gevc}$), the cross-section $\sigma_\mathrm{el}$ was obtained via linear interpolation between the Argonne values at \SIlist{5.0;6.0}{\gevc}~\cite{PhysRevD.9.1179}. The interpolation is performed bin-by-bin in $t$ and is considered a reasonable approximation since the cross-section dependence on the beam momentum is modest in this region. An exponential function was then fitted to the interpolated points and integrated over the $t$ normalization range, resulting in $\sigma_\mathrm{el} = \SI{2.46 \pm 0.02}{\milli\barn}$. \Cref{fig:luminosity-high-energy} shows the growth of the luminosity for the full beamtime period of 28 days. The recorded, time-integrated luminosity was found to be
\begin{equation*}
  \mathcal{L}=\SI[per-mode=power,uncertainty-descriptors={stat,norm,sys},separate-uncertainty-units=single]{5660.6 \pm 0.1 \pm 226 \pm 180}{\per\nano\barn}.
\end{equation*}
The first uncertainty is statistical. The second is systematic and comes from the normalization, resulting in an overall scale factor that is not related to HADES. The third is also systematic and comes from the HF selection criteria and integration range, as explained in \cref{sec:uncertainty_hf}.

\subsection{Time-integrated luminosity at $T = \SI{1.60 }{\gev}$}

At this energy, the normalization range was chosen to be $t = \brange{-1.38}{-0.42}{{\sgevc}}$. The cross section was obtained from the SAID fits (SM16 and WF16) of previously available data in the same range. This resulted in an integrated cross section of $\sigma_\mathrm{el}= \SI{2.5922 \pm 0.0015}{\milli\barn}$ in the normalization range. Using this value and the HH signal yield within the same range, the recorded, time-integrated luminosity at $T = \SI{1.60}{\gev}$ was determined to be 
\begin{equation*}
    \mathcal{L}=\SI[per-mode=power,uncertainty-descriptors={stat,norm,sys},separate-uncertainty-units=single]{345.1 \pm 0.1 \pm 2.7 \pm 2.1}{\per\nano\barn}.
\end{equation*}
The first uncertainty is statistical. The second is a systematic error arising from the normalization procedure. The third is a systematic error arising from the HH selection criteria, as summarized in \cref{sec:uncertainty_hh}.

\begin{figure}[!t]
 \centering
 \includegraphics[width=\linewidth]{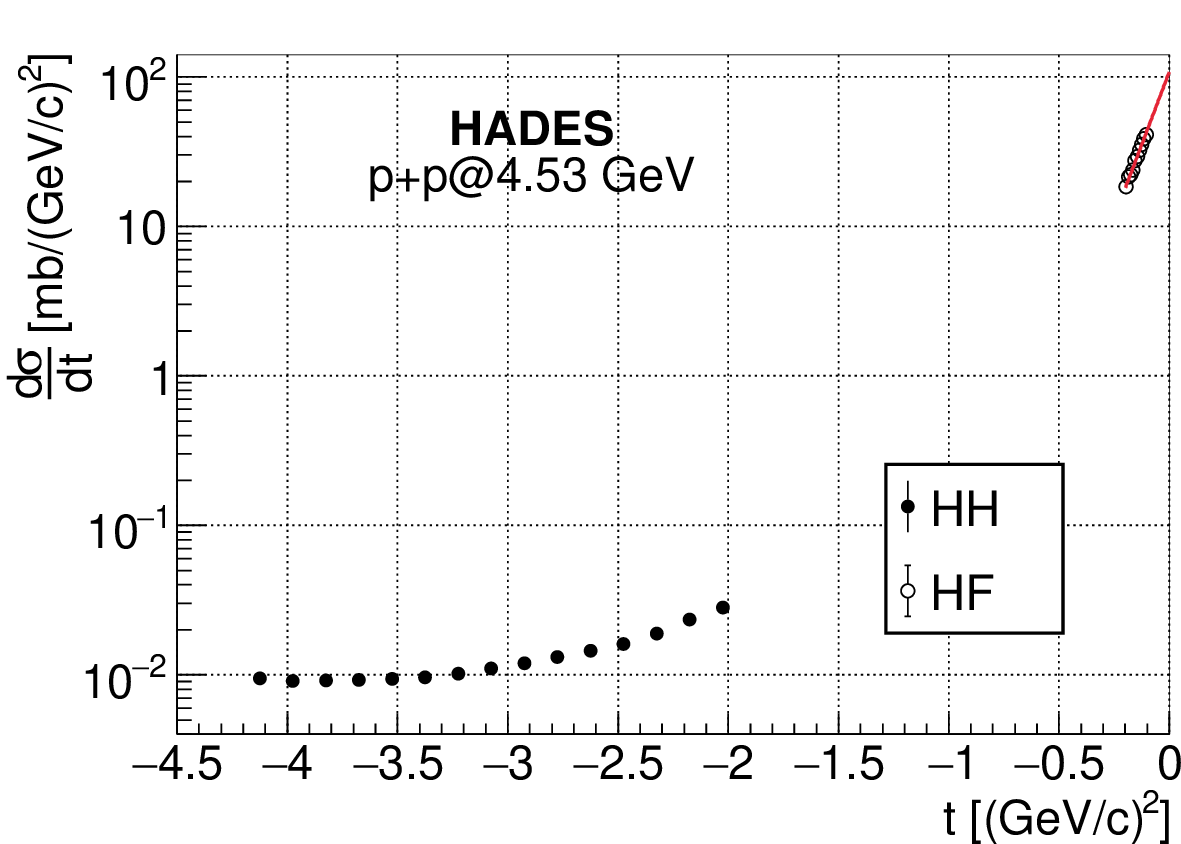}
 \caption{Combined HADES (HH+HF) $t$-dependent differential cross section for data collected at $T = \SI{4.53}{\gev}$. The uncertainties are smaller than the marker size. The red curve is a fit with \cref{eq:linear_exponential}.}
 \label{fig:combined-hades-highenergy}
\end{figure}

\section{Differential Cross Section}
\label{sec:crosssectionmeasurements}

\begin{figure}[!t]
 \centering
 \includegraphics[width=\linewidth]{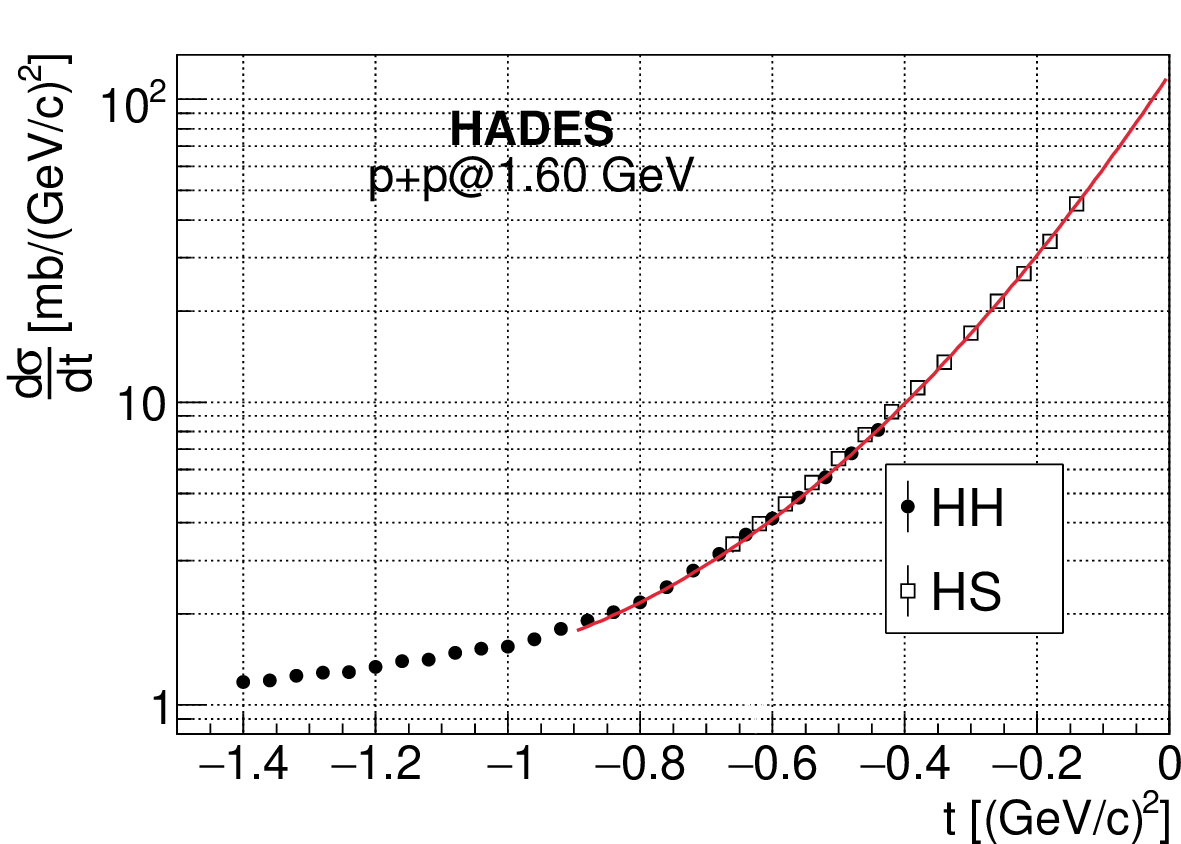}
 \caption{Combined HADES (HH+HS) $t$-dependent differential cross section for data collected at $T = \SI{1.60}{\gev}$. The uncertainties are smaller than the marker size. The red curve is a fit with \cref{eq:quadratic_exponential}.}
 \label{fig:combined-hades-lowenergy}
\end{figure}

The differential cross section $\dv*{\sigma}{t}$ in the $t$ regions with little or no coverage by previous experiments was obtained by normalizing the number of reconstructed and efficiency-corrected elastic-scattering events to the integrated luminosity determined for each beam energy. This means that \cref{eq:lumi-sum} is solved for the cross section using the recorded, time-integrated luminosity obtained in \cref{sec:luminosity}. \Cref{fig:combined-hades-highenergy,fig:combined-hades-lowenergy} show the resulting distributions as a function of $t$ for $T = \SI{4.53}{\gev}$ and $T = \SI{1.60}{\gev}$, respectively. The combination of the three independent analyses (HH, HS and HH) allows access to a wide $t$-range. At $T = \SI{4.53}{\gev}$ (\cref{fig:combined-hades-highenergy}), the spectrum is the result of the HH and HF analyses. The HS dataset, which would provide the overlap region between HH and HF, was excluded for this energy as explained in \cref{sec:eventselection}. At $T = \SI{1.60}{\gev}$ (\cref{fig:combined-hades-lowenergy}), the spectrum combines the HH and HS analyses. In this case, a clear overlap region in $t$ is covered by both datasets, showing the consistency of these independent analyses.  

\subsection{Comparison to World Data}
\label{subsec:comparisonworlddata}

\begin{figure}[!t]
    \centering
    \includegraphics[width=\linewidth]{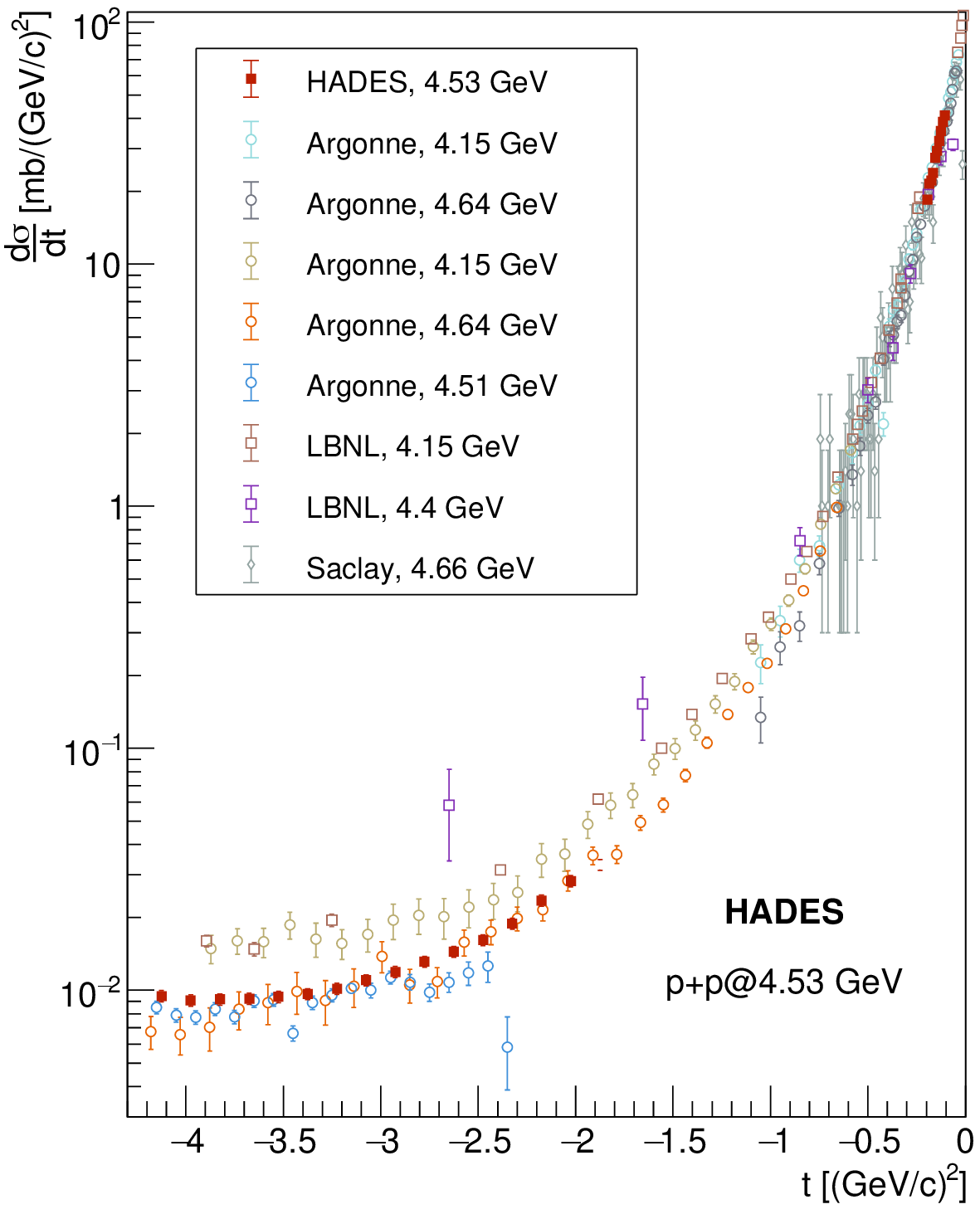}
    \caption{The pp elastic scattering differential cross section measured by HADES at $T = \SI{4.53}{\gev}$ compared to world data~\cite{PhysRevD.9.1179,PhysRevD.4.1309,Clyde:1966rta,PhysRev.170.1223, PhysRev.107.859,PhysRev.154.1284,PhysRevD.21.2445_BROO_low_energy}. The error bars of the HADES data are smaller than the marker size.}
    \label{fig:world-data-453}
\end{figure}

\begin{figure}[!t]
    \centering
    \includegraphics[width=\linewidth]{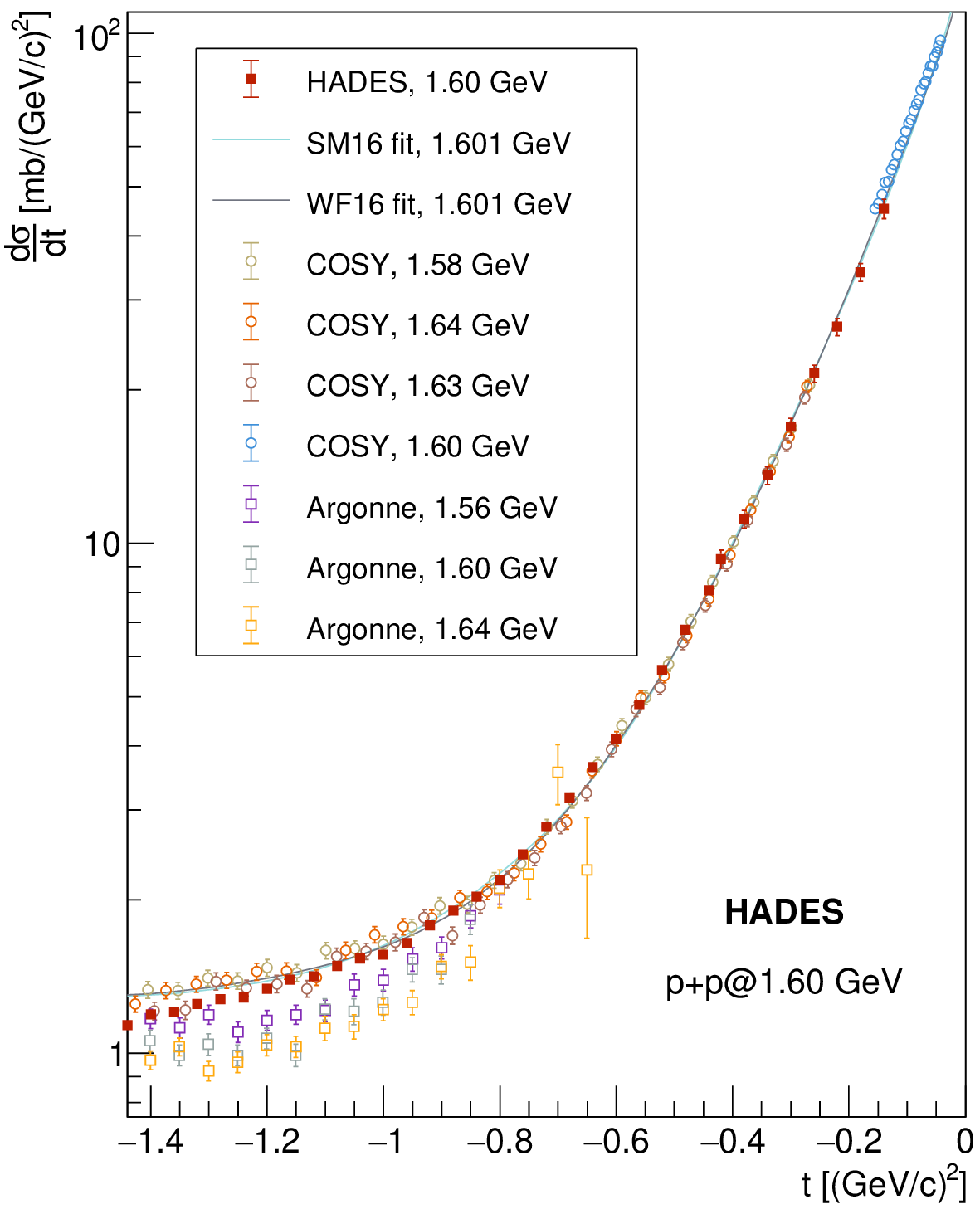}
    \caption{The pp elastic scattering differential cross section measured by HADES at $T = \SI{1.60}{\gev}$ compared to world data~\cite{EPJA.22.125_COSY_low_energy,PhysRevD.21.2445_BROO_low_energy,MCHEDLISHVILI201692_ANKE_pp_data} and the SAID fits SM16 and WF16~\cite{Workman:2016ysf}. The error bars of the HADES data are smaller than the marker size.}
    \label{fig:world-data-160}
\end{figure}

The differential cross section values determined here are presented together with data from Argonne, LBNL, Saclay and COSY in \cref{fig:world-data-453} ($T = \SI{4.53}{\gev}$) and \cref{fig:world-data-160} ($T = \SI{1.60}{\gev}$).

\subsubsection{Differential Cross Sections}
Previously published data closest in energy to the HADES measurement at $T = \SI{4.53}{\gev}$ are from Argonne~\cite{PhysRevD.9.1179,PhysRevD.4.1309,PhysRevD.21.2445_BROO_low_energy}, LBNL~\cite{Clyde:1966rta,PhysRev.170.1223, PhysRev.107.859} and Saclay~\cite{PhysRev.154.1284}. The new HADES data at small $|t|$ are normalized to the Argonne data. At large $|t|$, HADES provides an independent measurement of the differential cross section. The HADES data are in agreement with the Argonne measurements at similar energies and have a higher statistical precision.

At $T = \SI{1.60}{\gev}$, the world data are provided by EDDA~\cite{EPJA.22.125_COSY_low_energy} and Argonne~\cite{PhysRevD.21.2445_BROO_low_energy} in a large $|t|$ region and by ANKE~\cite{MCHEDLISHVILI201692_ANKE_pp_data} for small values of the momentum transfer. The HADES data are normalized with the SAID fit that is dominated by the EDDA data. The HH and HS data have been combined by calculating the uncertainty-weighted average in the overlapping $t$ region. The HADES data mostly agree with the SAID fit, also in $t$ regions that were not used in the normalization. A discrepancy can however be discerned for $t \leq 1.3 \mathrm{(GeV)}^2$. In this region, the SAID fit is dominated by the data from COSY. The COSY data disagree with those of Argonne, while the HADES data are between COSY and Argonne. The HADES measurement fills the gap in $t$ between the EDDA-at-COSY (green, orange and brown circles) and the ANKE-at-COSY (blue circles) measurements.

\subsubsection{Extraction of the Slope Parameter}
\label{subsec:fitfunction}

The data within the measured $|t|$ ranges were found to be well described by the functions given in \cref{eq:linear_exponential,eq:quadratic_exponential} at both $T = \SI{1.60}{\gev}$ and $T = \SI{4.53}{\gev}$. A fit of 
\cref{eq:linear_exponential}
to the data determines the nuclear slope parameter $B$. The optical point $\dv{\sigma_n}{t}|_{t=0}$ was a free parameter in the fit. However, since the optical point is strongly correlated with the normalization provided by the Argonne data and SAID, the obtained value is not considered an independent measurement.

At $T = \SI{4.53 }{\gev}$, the fit was performed using only the HF data, corresponding to the $t$ interval \brange{-0.2}{-0.1}{{\sgevc}}. The resulting slope parameter is $B = \SI[per-mode=power]{8.98 \pm 0.3}{(\gevc)^{-2}}$. The optical point was obtained from the fit to be $\dv{\sigma_n}{t}|_{t=0} = \SI{108.4 \pm 5.8}{\milli\barn\per{\sgevc}}$. The fit is shown in \cref{fig:combined-hades-highenergy}.

At $T = \SI{1.60 }{\gev}$, \cref{eq:quadratic_exponential} was fitted to the combined HH and HS data within a $t$ range of \brange{-0.9}{-0.14}{{\sgevc}} to determine the slope parameter $B$. This resulted in $B = \SI[per-mode=power]{7.5\pm0.3}{(\gevc)^{-2}}$, with an optical point of $\dv{\sigma_n}{t}|_{t=0} = \SI{121 \pm 10}{\milli\barn/{\sgevc}}$ and $C = \SI[per-mode=power]{3.1 \pm 0.2}{(\gevc)^{-4}}$. The fit is shown in \cref{fig:combined-hades-lowenergy}.

\begin{figure}[!t]
  \centering
  \includegraphics[width=1.0\linewidth]{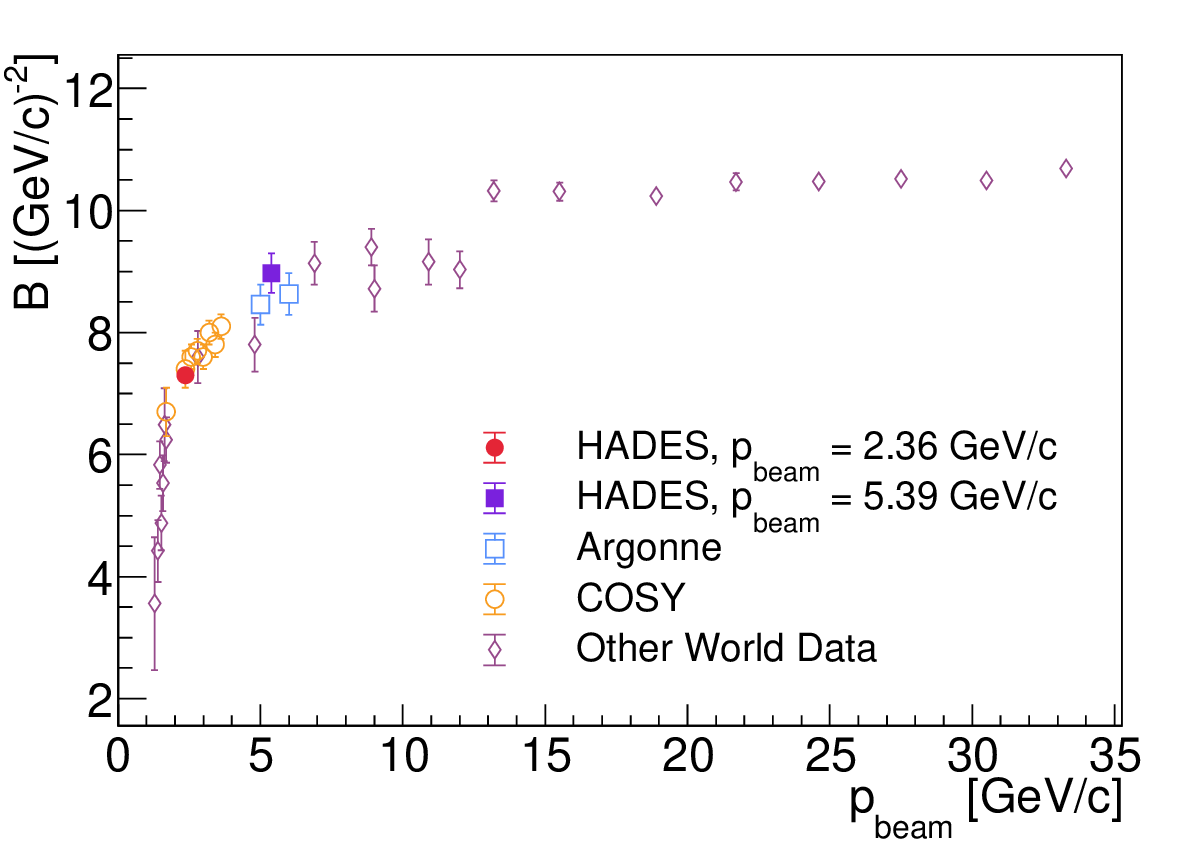}
  \caption{The slope parameter $B$ obtained from this work (filled points) and previous experiments at Argonne (empty squares), COSY (empty circles)~\cite{PhysRevD.9.1179} and World data (empty diamonds)~\cite{MCHEDLISHVILI201692_ANKE_pp_data,BEZNOGIKH197385,DOBROVOLSKY19831,PhysRevLett.31.1088,XU2021136022}.}
  \label{fig:slope-optical-parameter}
\end{figure}

The uncertainties of the obtained slope parameters are dominated by systematic errors. These were estimated by  varying the fit range and calculating the RMS of the resulting values with respect to the nominal range of the analysis. The HADES-specific systematic uncertainty but not the uncertainty arising from the normalization procedure are propagated to the uncertainty of the slope parameter $B$ and at $T = \SI{1.60}{\gev}$ also for the $C$ parameter. The slope parameters $B$ obtained at both beam momenta $p_{\mathrm{beam}}$ are shown to be consistent with the world data in \cref{fig:slope-optical-parameter}.

\section{Summary and Outlook}
\label{sec:discussion}

This work presents the first results from the upgraded HADES, including the new Forward Detector. The recorded, time-integrated luminosity has been measured for two beam energies by normalizing the signal yields in a $t$-region where overlapping data exists from previous experiments at Argonne and COSY. The measured time-integrated luminosities are now available for a large number of cross-sections of $\mathrm{pp}$-induced reactions~\cite{HADES:2020pcx}. This includes hyperon physics, di-lepton production, light mesons and baryonic resonances.

In addition, the $t$-dependent elastic \prot\prot\decays\prot\prot differential cross section has been measured in regions not covered by previous experiments. In particular, these data extend the range of elastic scattering measurements to low $|t|$-values. These new data can serve as a reference for future experiments by HADES and others. The $t$-dependent differential cross sections have also been used to extract the nuclear slope parameter. Our new HADES measurements follow the trend of previous measurements, while also providing unique information in regions where no data existed until now. Hence, the HADES data extend the database for elastic scattering and can be included in future partial wave analyses, \textit{e.g.} by SAID.

This report outlined and exploited methods to perform single-track momentum corrections and to study the beam alignment and beam energy. The studies presented in this paper provide fundamental input to all further analyses of data collected with this detector setup. For example, to fully benefit from techniques such as kinematic fits \cite{Esmail:2023yjg}, a good understanding of the beam-target system and the detector performance is essential.

\begin{acknowledgement}

The collaboration gratefully acknowledges the support by

\begin{sloppypar}
SIP JUC Cracow, Cracow (Poland), National Science Centre 2016/23/P/ST2/04066 POLONEZ, National Science Centre through grant SONATA-BIS no. 2023/50/E/ST2/00673, National Science Centre, Poland, 2017/26/M/ST2/00600; INP Cracow (Poland), National Science Centre grant nb. 2023/49/B/ST2/00652;WUT Warsaw (Poland) No: 2020/38/E/ST2/00019 (NCN), IDUB-POB-FWEiTE-3; TU Darmstadt, Darmstadt (Germany), VH-NG-823, DFG GRK 2128, DFG CRC-TR 211, BMBF:05P18RDFC1, HFHF (Campus Darmstadt), ELEMENTS 500/10.006, GSI F\&E, EMMI GSI Darmstadt; Goethe-University, Frankfurt (Germany), BMBF:05P12RFGHJ, GSI F\&E, HFHF (Campus Frankfurt), ELEMENTS 500/10.006; BU-Wuppertal, Wuppertal (Germany), BMBF 05P24PX1; JLU Giessen, Giessen (Germany), BMFTR 05P24RG6; IJCLab Orsay, Orsay (France), CNRS/IN2P3; NPI CAS, Rez, Rez (Czech Republic), MSMT LM2023060, MSMT OP JAK CZ.02.01.01/00/23$_{-}$015/0008181; The Swedish Research Council and the Knut and Alice Wallenberg foundation (Sweden). The authors thank the PANDA collaboration for the support with resources. This experiment was part of the FAIR-Phase0 program. The publication is funded by the OpenAccess Publishing Fund of GSI Helmholtzzentrum fuer Schwerionenforschung.
\end{sloppypar}

\end{acknowledgement}

\printbibliography

@string{EPJA = {Eur. Phys. J. A}}

@string{IJMPA = {Int. J. Mod. Phys. A}}

@string{JINST = {J. Instrum.}}

@string{JoPcs = {Jour. of Phys.: Conf. Ser.}}

@string{NIMA = {Nucl. Instrum. Meth. A}}

@string{NPA = {Nucl. Phys. A}}

@string{PLB = {Phys. Lett. B}}

@string{PR = {Phys. Rev.}}

@string{PRC = {Phys. Rev. C}}

@string{PRD = {Phys. Rev. D}}

@string{PRL = {Phys. Rev. Lett.}}

@string{PPNP = {Prog. Part. Nucl. Phys.}}

@article{Machleidt:2024bwl,
  author = "Machleidt, Ruprecht and Sammarruca, Francesca",
  title = "{Recent advances in chiral EFT based nuclear forces and their applications}",
  eprint = "2402.14032",
  archivePrefix = "arXiv",
  primaryClass = "nucl-th",
  doi = "10.1016/j.ppnp.2024.104117",
  journal = PPNP,
  volume = "137",
  pages = "104117",
  year = "2024"
}

@article{Islam:2005by,
  author = "Islam, M. M. and Luddy, R. J. and Prokudin, A. V.",
  title = "{Near forward $p p$ elastic scattering at LHC and nucleon structure}",
  eprint = "hep-ph/0508200",
  archivePrefix = "arXiv",
  doi = "10.1142/S0217751X06028473",
  journal = IJMPA,
  volume = "21",
  pages = "1--42",
  year = "2006"
}

@article{Selyugin:2012gv,
  author = "Selyugin, O. V.",
  title = "{Hard and Soft Pomerons in the Elastic Nucleon Scattering}",
  eprint = "1205.5867",
  archivePrefix = "arXiv",
  primaryClass = "hep-ph",
  doi = "10.1016/j.nuclphysa.2013.02.120",
  journal = NPA,
  volume = "903",
  pages = "54--64",
  year = "2013"
}

@article{Arndt:1997if,
  author = "Arndt, Richard A. and Oh, Chang Heon and Strakovsky, Igor I. and Workman, Ron L. and Dohrmann, Frank",
  title = "{Nucleon-nucleon elastic scattering analysis to 2.5-GeV}",
  eprint = "nucl-th/9706003",
  archivePrefix = "arXiv",
  reportNumber = "VPI-CAPS-97-6",
  doi = "10.1103/PhysRevC.56.3005",
  journal = PRC,
  volume = "56",
  pages = "3005--3013",
  year = "1997"
}

@article{Shirokov,
  title = {Properties of nuclear matter within the JISP16 $NN$ interaction},
  author = {Shirokov, A. M. and Negoita, A. G. and Vary, J. P. and Bogner, S. K. and Mazur, A. I. and Mazur, E. A. and Gogny, D.},
  journal = PRC,
  volume = {90},
  issue = {2},
  pages = {024324},
  numpages = {5},
  year = {2014},
  publisher = {American Physical Society},
  doi = {10.1103/PhysRevC.90.024324},
  url = {https://link.aps.org/doi/10.1103/PhysRevC.90.024324}
}

@article{Dymov:2015hku,
  author = "Dymov, S. and others",
  title = "{Analysing powers and spin correlations in deuteron-proton charge exchange at 726 MeV}",
  eprint = "1503.00514",
  archivePrefix = "arXiv",
  primaryClass = "nucl-ex",
  doi = "10.1016/j.physletb.2015.04.019",
  journal = PLB,
  volume = "744",
  pages = "391--394",
  year = "2015"
}

@article{Workman:2016ysf,
  author = "Workman, Ron L. and Briscoe, William J. and Strakovsky, Igor I.",
  title = "{Partial-Wave Analysis of Nucleon-Nucleon Elastic Scattering Data}",
  eprint = "1609.01741",
  archivePrefix = "arXiv",
  primaryClass = "nucl-th",
  doi = {10.1103/PhysRevC.94.065203},
  journal = PRC,
  volume = {94},
  number = {6},
  pages = {065203},
  year = {2016}
}

@article{Uzhinsky:2016uhg,
    author = "Uzhinsky, V. and Galoyan, A. and Hu, Q. and Ritman, J. and Xu, H.",
    title = "{Empirical parametrization of the nucleon-nucleon elastic scattering amplitude at high beam momenta for Glauber calculations and Monte Carlo simulations}",
    eprint = "1603.04731",
    archivePrefix = "arXiv",
    primaryClass = "hep-ph",
    doi = "10.1103/PhysRevC.94.064003",
    journal = "Phys. Rev. C",
    volume = "94",
    number = "6",
    pages = "064003",
    year = "2016"
}

@misc{SAID,
  author = {SAID},
  url = {https://gwdac.phys.gwu.edu/analysis/nn_analysis.html},
  note = {The George Washington University, INS Data Analysis Center, accessed 2023}
}

@article{PhysRevD.4.1309,
  title = {Large-Angle Proton-Proton Elastic Scattering at Intermediate Momenta},
  author = {Kammerud, R. C. and Brabson, B. B. and Crittenden, R. R. and Heinz, R. M. and Neal, H. A. and Paik, H. W. and Sidwell, R. A.},
  journal = PRD,
  volume = {4},
  issue = {5},
  pages = {1309--1324},
  year = {1971},
  publisher = {American Physical Society},
  doi = {10.1103/PhysRevD.4.1309},
  url = {https://link.aps.org/doi/10.1103/PhysRevD.4.1309}
}

@article{PhysRev.154.1284,
  title = {Proton-Proton Interactions at 5.5 GeV/c},
  author = {Alexander, G. and Benary, O. and Czapek, G. and Haber, B. and Kidron, N. and Reuter, B. and Shapira, A. and Simopoulou, E. and Yekutieli, G.},
  journal = PR,
  volume = {154},
  issue = {5},
  pages = {1284--1304},
  numpages = {0},
  year = {1967},
  publisher = {American Physical Society},
  doi = {10.1103/PhysRev.154.1284},
  url = {https://link.aps.org/doi/10.1103/PhysRev.154.1284}
}

@article{PhysRev.107.859,
  title = {Elastic Proton-Proton Scattering at 2.24, 4.40, and 6.15 Bev},
  author = {Cork, Bruce and Wenzel, William A. and Causey, Charles W.},
  journal = PR,
  volume = {107},
  issue = {3},
  pages = {859--867},
  numpages = {0},
  year = {1957},
  publisher = {American Physical Society},
  doi = {10.1103/PhysRev.107.859},
  url = {https://link.aps.org/doi/10.1103/PhysRev.107.859}
}

@article{PhysRev.170.1223,
  title = {Nucleon Isobar Production in Proton-Proton Collisions between 3 and 7 $\frac{\mathrm{GeV}}{c}$},
  author = {Ankenbrandt, C. M. and Clark, A. R. and Cork, Bruce and Elioff, T. and Kerth, L. T. and Wenzel, W. A.},
  journal = PR,
  volume = {170},
  issue = {5},
  pages = {1223--1236},
  numpages = {0},
  year = {1968},
  publisher = {American Physical Society},
  doi = {10.1103/PhysRev.170.1223},
  url = {https://link.aps.org/doi/10.1103/PhysRev.170.1223}
}

@phdthesis{Clyde:1966rta,
  author = "Clyde, A. R.",
  title = "{Proton-proton elastic scattering at incidence momenta 3, 5, and 7 BeV/c}",
  reportNumber = "UCRL-16275",
  school = "Calif. U. Berkeley",
  year = "1966"
}

@article{PhysRevD.9.1179,
  title = {Systematic study of ${\ensuremath{\pi}}^{\ifmmode\pm\else\textpm\fi{}}p$, ${K}^{\ifmmode\pm\else\textpm\fi{}}p$, $\mathrm{pp}$, and $\overline{p}p$ forward elastic scattering from $3 \mathrm{to} 6 \frac{\mathrm{GeV}}{c}$},
  author = {Ambats, I. and Ayres, D. S. and Diebold, R. and Greene, A. F. and Kramer, S. L. and Lesnik, A. and Rust, D. R. and Ward, C. E. W. and Wicklund, A. B. and Yovanovitch, D. D.},
  journal = PRD,
  volume = {9},
  issue = {5},
  pages = {1179--1209},
  numpages = {0},
  year = {1974},
  publisher = {American Physical Society},
  doi = {10.1103/PhysRevD.9.1179},
  url = {https://link.aps.org/doi/10.1103/PhysRevD.9.1179}
}

@article{EPJA.22.125_COSY_low_energy,
  title = {A precision measurement of pp elastic scattering cross-sections
  at intermediate energies},
  author = {{The EDDA Collaboration}},
  journal = EPJA,
  volume = {22},
  pages = {125-148},
  year = {2004},
  doi = {10.1140/epja/i2004-10011-3}
}

@article{PhysRevD.21.2445_BROO_low_energy,
  title = {Measurement of wide-angle elastic scattering of pions and protons off protons},
  author = {Jenkins, K. A. and Price, L. E. and Klem, R. and Miller, R. J. and Schreiner, P. and Marshak, M. L. and Peterson, E. A. and Ruddick, K.},
  journal = PRD,
  volume = {21},
  issue = {9},
  pages = {2445--2496},
  numpages = {0},
  year = {1980},
  publisher = {American Physical Society},
  doi = {10.1103/PhysRevD.21.2445},
  url = {https://link.aps.org/doi/10.1103/PhysRevD.21.2445}
}

@article{MCHEDLISHVILI201692_ANKE_pp_data,
  title = {Measurement of the absolute differential cross section of proton--proton elastic scattering at small angles},
  journal = PLB,
  volume = {755},
  pages = {92-96},
  year = {2016},
  issn = {0370-2693},
  doi = {https://doi.org/10.1016/j.physletb.2016.01.066},
  url = {https://www.sciencedirect.com/science/article/pii/S037026931600085X},
  author = {D. Mchedlishvili and D. Chiladze and S. Dymov and Z. Bagdasarian and S. Barsov and R. Gebel and B. Gou and M. Hartmann and A. Kacharava and I. Keshelashvili and A. Khoukaz and P. Kulessa and A. Kulikov and A. Lehrach and N. Lomidze and B. Lorentz and R. Maier and G. Macharashvili and S. Merzliakov and S. Mikirtychyants and M. Nioradze and H. Ohm and D. Prasuhn and F. Rathmann and V. Serdyuk and D. Schroer and V. Shmakova and R. Stassen and H.J. Stein and H. Stockhorst and I.I. Strakovsky and H. Ströher and M. Tabidze and A. Täschner and S. Trusov and D. Tsirkov and Yu. Uzikov and Yu. Valdau and C. Wilkin and R.L. Workman and P. Wüstner}
}

@article{Agakichiev_2009_HADES,
  doi = {10.1140/epja/i2009-10807-5},
  url = {https://doi.org/10.1140%2Fepja%2Fi2009-10807-5},
  year = 2009,
  publisher = {Springer Science and Business Media {LLC}},
  volume = {41},
  number = {2},
  pages = {243--277},
  author = {G. Agakichiev and others},
  title = {The high-acceptance dielectron spectrometer {HADES}},
  journal = EPJA
}

@article{HADES:2020pcx_forward_detector,
  author = "Adamczewski-Musch, J. and others",
  collaboration = "HADES, PANDA",
  title = "{Production and electromagnetic decay of hyperons: a feasibility study with HADES as a phase-0 experiment at FAIR}",
  eprint = "2010.06961",
  archivePrefix = "arXiv",
  primaryClass = "nucl-ex",
  doi = "10.1140/epja/s10050-021-00388-w",
  journal = EPJA,
  volume = "57",
  number = "4",
  pages = "138",
  year = "2021"
}

@article{GRZONKA2022167410,
  title = {A large area efficient trigger scintillator with SiPM read out},
  journal = NIMA,
  volume = {1041},
  pages = {167410},
  year = {2022},
  issn = {0168-9002},
  doi = {https://doi.org/10.1016/j.nima.2022.167410},
  url = {https://www.sciencedirect.com/science/article/pii/S0168900222007021},
  author = {D. Grzonka and P. Bergmann and T. {Hahnraths von der Gracht} and P. Kulessa and W. Parol and T. Sefzick and J. Ritman and M. Zielinski}
}

@article{Smyrski:2017,
  author = {Smyrski, J. and Apostolou, A. and Biernat, J. and Czyżycki, W. and Filo, G. and Fioravanti, E. and Fiutowski, T. and Gianotti, P. and Idzik, M. and Korcyl, G. and Korcyl, K. and Lisowski, E. and Lisowski, F. and Płażek, J. and Przyborowski, D. and Przygoda, W. and Ritman, J. and Salabura, P. and Savrie, M. and Strzempek, P. and Swientek, K. and Wintz, P. and Wrońska, A.},
  doi = {10.1088/1748-0221/12/06/c06032},
  journal = JINST,
  number = {06},
  pages = {C06032},
  publisher = {{IOP} Publishing},
  title = {Design of the forward straw tube tracker for the {PANDA} experiment},
  url = {https://doi.org/10.1088%2F1748-0221%2F12%2F06%2Fc06032},
  volume = {12},
  year = 2017
}

@article{Smyrski:2018,
  author = {Smyrski, J. and Fiutowski, T. and Gianotti, P. and Heczko, A. and Idzik, M. and Kajetanowicz, M. and Korcyl, G. and Korzeniak, B. and Lalik, R. and Lisowski, E. and Malarz, A. and Migdał, W. and Misiak, A. and Przygoda, W. and Ritman, J. and Salabura, P. and Savrie, M. and Swientek, K. and Wintz, P. and Wrońska, A.},
  doi = {10.1088/1748-0221/13/06/p06009},
  journal = JINST,
  number = {06},
  pages = {P06009},
  publisher = {{IOP} Publishing},
  title = {Pressure stabilized straw tube modules for the {PANDA} Forward Tracker},
  url = {https://doi.org/10.1088%2F1748-0221%2F13%2F06%2Fp06009},
  volume = {13},
  year = 2018
}

@article{Wintz:2014uwa,
  author = {Wintz, Peter},
  collaboration = {Panda Tracking Group},
  doi = {10.1007/s10751-014-1035-6},
  journal = {Hyperfine Interactions},
  number = {1-3},
  pages = {147–152},
  title = {{The central straw tube tracker in the $\bar {P}ANDA$ experiment}},
  volume = {229},
  year = 2014,
  %%editor = {Johansson, T. and Froehlich, P. and Jonsell, S.}
}

@article{FINCK200363,
  author = {Finck, Ch. and Fonte, P. and Gobbi, A.},
  doi = {10.1016/S0168-9002(03)01278-6},
  issn = {0168-9002},
  journal = NIMA,
  note = {Proceedings of the 6th Int. Workshop on Resistive Plate Chambers and Related Detectors},
  number = {1},
  pages = {63–69},
  title = {Results concerning understanding and applications of timing GRPCs},
  url = {http://www.sciencedirect.com/science/article/pii/S0168900203012786},
  volume = {508},
  year = {2003}
}

@article{pluto,
  doi = {10.1088/1742-6596/219/3/032039},
  url = {https://dx.doi.org/10.1088/1742-6596/219/3/032039},
  year = {2010},
  publisher = {},
  volume = {219},
  number = {3},
  pages = {032039},
  author = {I Fröhlich and  T Galatyuk and  R Holzmann and  J Markert and  B Ramstein and  P Salabura and  J Stroth},
  title = {Design of the pluto event generator},
  journal = JoPcs
}

@article{KRUGER2022167046,
  title = {LGAD technology for HADES, accelerator and medical applications},
  journal = NIMA,
  volume = {1039},
  pages = {167046},
  year = {2022},
  issn = {0168-9002},
  doi = {https://doi.org/10.1016/j.nima.2022.167046},
  url = {https://www.sciencedirect.com/science/article/pii/S0168900222004697},
  author = {W. Krüger and T. Bergauer and T. Galatyuk and A. Hirtl and V. Kedych and M. Kis and S. Linev and J. Michel and J. Pietraszko and F. Pitters and A. Rost and C.J. Schmidt and V. Svintozelskyi and M. Träger and M. Traxler and F. Ulrich-Pur and Ch. Wendisch}
}

@article{Esmail:2023yjg,
  author = {Esmail, Waleed and Rieger, Jana and Taylor, Jenny and Bohman, Malin and Sch{\"o}nning, Karin},
  title = "{KinFit: A Kinematic Fitting Package for Hadron Physics Experiments}",
  eprint = "2308.09575",
  archivePrefix = "arXiv",
  primaryClass = "physics.data-an",
  doi = "10.1007/s41781-023-00112-x",
  journal = "Comput. Softw. Big Sci.",
  volume = "8",
  number = "1",
  pages = "3",
  year = "2024"
}

@article{PANDA:2021ozp,
    author = "Barucca, G. and others",
    collaboration = "PANDA",
    title = "{PANDA Phase One}",
    eprint = "2101.11877",
    archivePrefix = "arXiv",
    primaryClass = "hep-ex",
    doi = "10.1140/epja/s10050-021-00475-y",
    journal = EPJA,
    volume = "57",
    number = "6",
    pages = "184",
    year = "2021"
}

@article{HADES:2020pcx,
    author = "Adamczewski-Musch, J. and others",
    collaboration = "HADES, PANDA",
    title = "{Production and electromagnetic decay of hyperons: a feasibility study with HADES as a phase-0 experiment at FAIR}",
    eprint = "2010.06961",
    archivePrefix = "arXiv",
    primaryClass = "nucl-ex",
    doi = "10.1140/epja/s10050-021-00388-w",
    journal = EPJA,
    volume = "57",
    number = "4",
    pages = "138",
    year = "2021"
}

@article{PECHENOVA201540,
title = {The alignment strategy of HADES},
journal = NIMA,
volume = {785},
pages = {40-46},
year = {2015},
issn = {0168-9002},
doi = {https://doi.org/10.1016/j.nima.2015.03.003},
url = {https://www.sciencedirect.com/science/article/pii/S0168900215002934},
author = {O. Pechenova and V. Pechenov and T. Galatyuk and T. Hennino and R. Holzmann and G. Kornakov and J. Markert and C. Müntz and P. Salabura and A. Schmah and E. Schwab and J. Stroth},
}

@article{BEZNOGIKH197385,
title = {Total elastic p-p, p-d, p-n cross sections in the energy range of 1-70 GeV},
author = {G.G. Beznogikh and A. Bujak and V.A. Nikitin and M.G. Shafranova and V.A. Sviridov and Truong Bien and L.V. Vikhlyantseva and V.I. Zayachki and L.S. Zolin},
journal = PLB,
volume = {43},
number = {1},
pages = {85-88},
year = {1973},
issn = {0370-2693},
doi = {https://doi.org/10.1016/0370-2693(73)90550-9},
url = {https://www.sciencedirect.com/science/article/pii/0370269373905509}
}

@article{DOBROVOLSKY19831,
title = {Small angle pp scattering at energies from 650 to 1000 MeV},
author = {A.V. Dobrovolsky and A.V. Khanzadeev and G.A. Korolev and E.M. Maev and V.I. Medvedev and G.L. Sokolov and N.K. Terentyev and Y. Terrien and G.N. Velichko and A.A. Vorobyov and Yu.K. Zalite},
journal = {Nucl. Phys. B},
volume = {214},
number = {1},
pages = {1-20},
year = {1983},
issn = {0550-3213},
doi = {https://doi.org/10.1016/0550-3213(83)90163-3},
url = {https://www.sciencedirect.com/science/article/pii/0550321383901633},
}

@article{PhysRevLett.31.1088,
  title = {Measurement of the Slope of the Diffraction Peak for Elastic $p\ensuremath{-}p$ Scattering from 8 to 400 GeV},
  author = {Bartenev, V. and Kuznetsov, A. and Morozov, B. and Nikitin, V. and Pilipenko, Y. and Popov, V. and Zolin, L. and Carrigan, R. A. and Malamud, E. and Yamada, R. and Cool, R. L. and Goulianos, K. and Chiang, I -Hung and Melissinos, A. C. and Gross, D. and Olsen, S. L.},
  journal = PRL,
  volume = {31},
  issue = {17},
  pages = {1088--1091},
  numpages = {0},
  year = {1973},
  publisher = {American Physical Society},
  doi = {10.1103/PhysRevLett.31.1088},
  url = {https://link.aps.org/doi/10.1103/PhysRevLett.31.1088}
}

@article{XU2021136022,
title = {Measurement of proton-proton elastic scattering into the Coulomb region at Pbeam = 2.5, 2.8 and 3.2 GeV/c},
author = {H. Xu and Y. Zhou and U. Bechstedt and J. Böker and A. Gillitzer and F. Goldenbaum and D. Grzonka and Q. Hu and A. Khoukaz and F. Klehr and B. Lorentz and D. Prasuhn and J. Ritman and S. Schadmand and T. Sefzick and T. Stockmanns and I.I. Strakovsky and A. Täschner and C. Wilkin and R.L. Workman and P. Wüstner},
journal = PLB,
volume = {812},
pages = {136022},
year = {2021},
issn = {0370-2693},
doi = {https://doi.org/10.1016/j.physletb.2020.136022},
url = {https://www.sciencedirect.com/science/article/pii/S037026932030825X},
}

@article{BLANCO2023167652,
  author = {Blanco, A. and Fonte, P. and Lopes, L. and Saraiva, J.},
  doi = {10.1016/j.nima.2022.167652},
  issn = {0168-9002},
  journal = NIMA,
  pages = {167652},
  title = {{Improving count rate capability of timing RPCs by increasing the detector working temperature}},
  url = {https://www.sciencedirect.com/science/article/pii/S0168900222009445},
  volume = {1045},
  year = {2023}
}

@article{Goslawski:2009vf,
    author = "Goslawski, P. and others",
    title = "{High precision beam momentum determination in a synchrotron using a spin resonance method}",
    eprint = "0908.3103",
    archivePrefix = "arXiv",
    primaryClass = "physics.acc-ph",
    doi = "10.1103/PhysRevSTAB.13.022803",
    journal = "Phys. Rev. ST Accel. Beams",
    volume = "13",
    pages = "022803",
    year = "2010"
}

\clearpage
\appendix
\section*{Appendix}

\begin{table*}[!ht]
  \caption{Data points at $T = \SI{4.53}{\gev}$.}
  \label{tab:data_points_453}
  \centering
  \begin{tabular}{SSSSS}
  \toprule
    $t$ [\si{{\sgevc}}] &
    $\dv{\sigma}{t}$ [\si{\milli\barn\per{\sgevc}}] &
    $\sigma_\mathrm{stat}$ [\si{\milli\barn\per{\sgevc}}] &
    $\sigma_\mathrm{sys}$ [\si{\milli\barn\per{\sgevc}}] &
    $\sigma_\mathrm{norm}$ [\si{\milli\barn\per{\sgevc}}] \\
  \midrule
  \multicolumn{5}{c}{HH} \\
  \midrule
-4.125  & 0.00947   &  0.00004  &  0.00014  &  0.0005 \\ 
-3.975  & 0.00909   &  0.00004  &  0.00014  &  0.0005 \\ 
-3.825  & 0.00917   &  0.00004  &  0.00014  &  0.0005 \\ 
-3.675  & 0.00925   &  0.00004  &  0.00014  &  0.0005 \\ 
-3.525  & 0.00940   &  0.00004  &  0.00014  &  0.0005 \\ 
-3.375  & 0.00964   &  0.00004  &  0.00015  &  0.0005 \\ 
-3.225  & 0.01015  &  0.00004  &  0.00016  &  0.0005 \\ 
-3.075  & 0.01100  &  0.00004  &  0.00017  &  0.0006 \\ 
-2.925  & 0.01192  &  0.00005  &  0.00018  &  0.0006 \\ 
-2.775  & 0.01314  &  0.00005  &  0.00020  &  0.0007 \\ 
-2.625  & 0.01444  &  0.00005  &  0.00022  &  0.0007 \\ 
-2.475  & 0.01612  &  0.00005  &  0.00025  &  0.0008 \\ 
-2.325  & 0.0189   &  0.00006  &  0.00029  &  0.0010 \\ 
-2.175  & 0.0235   &  0.00007  &  0.00036  &  0.0012 \\ 
-2.025  & 0.0283   &  0.00007  &  0.00043  &  0.0014 \\ 
  \midrule
  \multicolumn{5}{c}{HF} \\
  \midrule
-0.1950  & 18.5  &  0.007   &  0.6  &  0.7 \\
-0.1850  & 21.5  &  0.007   &  0.7  &  0.9 \\
-0.1750  & 22.2  &  0.008   &  0.7  &  0.9 \\
-0.1650  & 23.8  &  0.009   &  0.8  &  1.0 \\
-0.1550  & 27.5  &  0.011   &  0.9  &  1.1 \\
-0.1450  & 29.4  &  0.012   &  0.9  &  1.2 \\
-0.1350  & 32.3  &  0.015   &  1.0  &  1.3 \\
-0.1250  & 35.4  &  0.020   &  1.1  &  1.4 \\
-0.1150  & 38.8  &  0.026   &  1.2  &  1.6 \\
-0.1050  & 41.2  &  0.0352  &  1.3  &  1.6 \\
  \bottomrule
  \end{tabular}
\end{table*}

\begin{table*}[!ht]
  \caption{Data points at $T = \SI{1.60}{\gev}$.}
  \label{tab:data_points_160}
  \centering
  \begin{tabular}{
  S[table-format=+2.3]
  SSSS}
    \toprule
    $t$ [\si{{\sgevc}}] &
    $\dv{\sigma}{t}$ [\si{\milli\barn\per{\sgevc}}] &
    $\sigma_\mathrm{stat}$ [\si{\milli\barn\per{\sgevc}}] &
    $\sigma_\mathrm{sys}$ [\si{\milli\barn\per{\sgevc}}] &
    $\sigma_\mathrm{norm}$ [\si{\milli\barn\per{\sgevc}}] \\
    \midrule
-1.440  & 1.136  &  0.0030  &  0.007    &  0.009 \\ 
-1.400  & 1.191  &  0.0032  &  0.007    &  0.010 \\ 
-1.360  & 1.204  &  0.0032  &  0.007    &  0.010 \\ 
-1.320  & 1.249  &  0.0033  &  0.008    &  0.010 \\ 
-1.280  & 1.277  &  0.0034  &  0.008    &  0.010 \\ 
-1.240  & 1.286  &  0.0033  &  0.008    &  0.010 \\ 
-1.200  & 1.337  &  0.0034  &  0.008    &  0.011 \\ 
-1.160  & 1.394  &  0.0034  &  0.009    &  0.011 \\ 
-1.120  & 1.414  &  0.0034  &  0.009    &  0.011 \\ 
-1.080  & 1.484  &  0.0035  &  0.009    &  0.012 \\ 
-1.040  & 1.532  &  0.004    &  0.009    &  0.012 \\ 
-1.000  & 1.560  &  0.004    &  0.010  &  0.012 \\ 
-0.9600  & 1.646  &  0.004    &  0.010  &  0.013 \\ 
-0.9200  & 1.781  &  0.004    &  0.011  &  0.014 \\ 
-0.8800  & 1.903  &  0.004    &  0.012  &  0.015 \\ 
-0.8400  & 2.026  &  0.004    &  0.012  &  0.016 \\ 
-0.8000  & 2.180  &  0.005    &  0.013  &  0.017 \\ 
-0.7600  & 2.452  &  0.005    &  0.015  &  0.020 \\ 
-0.7200  & 2.778  &  0.005    &  0.017  &  0.022 \\ 
-0.6800  & 3.160  &  0.006    &  0.019  &  0.025 \\ 
-0.6406  & 3.640  &  0.004    &  0.022  &  0.022 \\ 
-0.6008  & 4.130  &  0.004    &  0.025  &  0.025 \\ 
-0.5608  & 4.822  &  0.004    &  0.029  &  0.030 \\ 
-0.5209  & 5.646  &  0.005    &  0.034  &  0.035 \\ 
-0.4811  & 6.76   &  0.005    &  0.04    &  0.04 \\ 
-0.4412  & 8.08   &  0.006    &  0.05    &  0.05 \\ 
-0.4200  & 9.31   &  0.0031  &  0.4    &  0.09 \\ 
-0.3800  & 11.2   &  0.0031  &  0.4    &  0.11 \\ 
-0.3400  & 13.6   &  0.0033  &  0.5    &  0.14 \\ 
-0.3000  & 16.9   &  0.0034  &  0.6    &  0.17 \\ 
-0.2600  & 21.5   &  0.0035  &  0.8    &  0.22 \\ 
-0.2200  & 26.6   &  0.0035  &  1.0   &  0.27 \\ 
-0.1800  & 34.0   &  0.0035  &  1.3  &  0.34 \\ 
-0.1400  & 45.3   &  0.0036  &  2.0  &  0.46 \\ 
  \bottomrule
  \end{tabular}
\end{table*}
\end{document}